\documentclass[10pt,twocolumn,letterpaper]{article}

\usepackage{iccv}  

\usepackage[accsupp]{axessibility} 
\usepackage{bm}
\usepackage{bbm}

\usepackage{colortbl}
\usepackage{graphicx}
\usepackage{tabularx}
\usepackage{makecell}
\usepackage{subcaption}
\usepackage{multirow}
\usepackage{amsmath}
\usepackage{mathrsfs}

\usepackage{xspace}

\newcommand{\keep}[1]{}
\newcommand{\old}[1]{}

\usepackage{url}

\definecolor{iccvblue}{rgb}{0.21,0.49,0.74}
\usepackage[pagebackref,breaklinks,colorlinks,allcolors=iccvblue]{hyperref}

\def\paperID{3694} 
\def\confName{ICCV}
\def\confYear{2025}

\title{
GestureHYDRA: Semantic Co-speech Gesture Synthesis 
\\  via Hybrid Modality Diffusion Transformer and
Cascaded-Synchronized Retrieval-Augmented Generation
}

\author{
Quanwei Yang$^{1*}$ \quad
Luying Huang$^{2*}$ \quad
Kaisiyuan Wang$^{2\dagger}$ \quad
Jiazhi Guan$^{2}$ \quad
Shengyi He$^{2}$ \quad
Fengguo Li$^{2}$ \\
Hang Zhou$^{2}$ \quad
Lingyun Yu$^{1}$ \quad
Yingying Li$^{2}$ \quad
Haocheng Feng$^{2}$ \quad
Hongtao Xie$^{1\dagger}$ \\
$^{1}$University of Science and Technology of China \quad
$^{2}$Baidu Inc. \\
\small $^*$Equal contribution. \quad $^\dagger$Corresponding authors.
}

\begin{document}

\twocolumn[{
\renewcommand\twocolumn[1][]{#1}%
\maketitle
\begin{center}
 \centering
 \includegraphics[width=\textwidth]{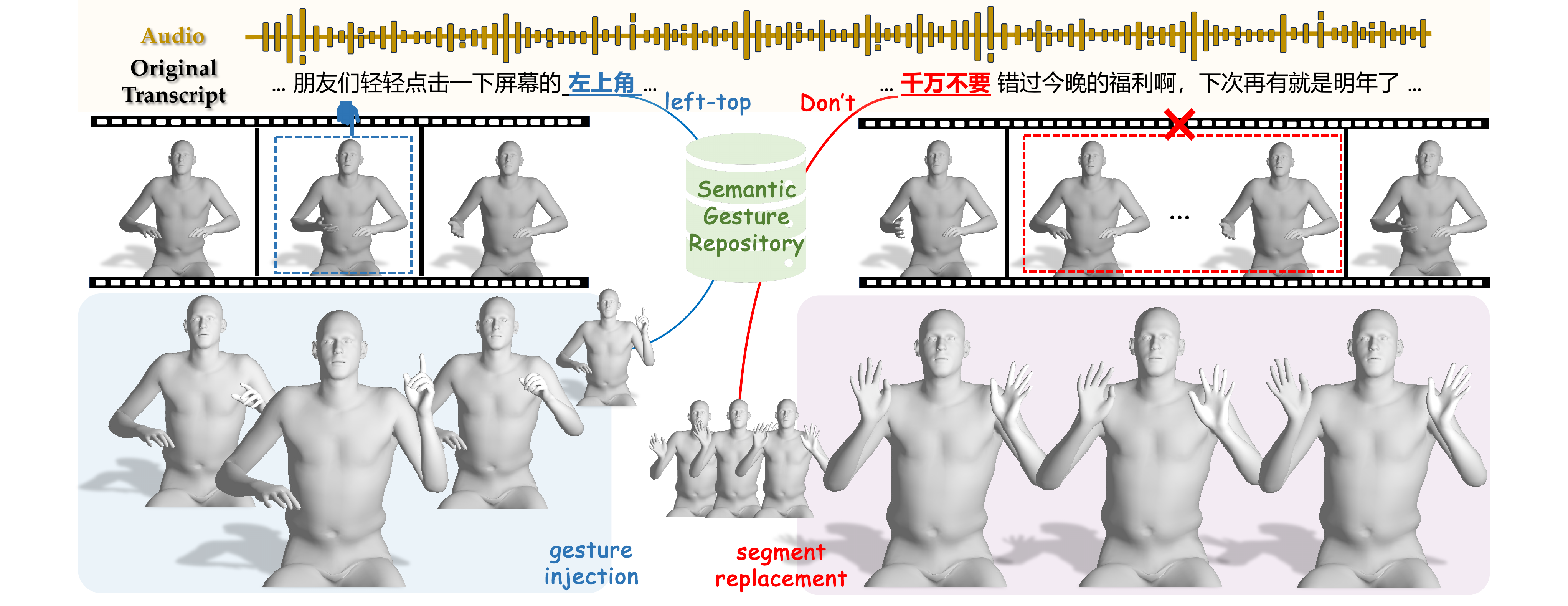}
\captionof{figure}{
We propose \textbf{GestureHYDRA}, a co-speech gesture generation system based on
a hybrid-modality diffusion transformer and a cascaded-synchronized Retrieval-Augmented Generation to enable flexible and reliable human gesture synthesis.
}
\label{fig:teasor}
\end{center}
}]

\begin{abstract}
While increasing attention has been paid to co-speech gesture synthesis, most previous works neglect to investigate hand gestures with explicit and essential semantics.
In this paper, we study co-speech gesture generation with an emphasis on specific hand gesture activation, which can deliver more instructional information than common body movements.
To achieve this, we first build a high-quality dataset of 3D human body movements including a set of semantically explicit hand gestures that are commonly used by live streamers.
Then we present a hybrid-modality gesture generation system GestureHYDRA built upon a hybrid-modality diffusion transformer architecture with novelly designed motion-style injective transformer layers, which enables advanced gesture modeling ability and versatile gesture operations.
To guarantee these specific hand gestures can be activated, we introduce a cascaded retrieval-augmented generation strategy built upon a semantic gesture repository annotated for each subject and an adaptive audio-gesture synchronization mechanism, which substantially improves semantic gesture activation and production efficiency.
Quantitative and qualitative experiments demonstrate that our proposed approach achieves superior performance over all the counterparts.
The project page can be found at \href{https://mumuwei.github.io/GestureHYDRA/}{\textcolor{red}{https://mumuwei.github.io/GestureHYDRA/}}.

\end{abstract}

\section{Introduction}
Co-speech gesture synthesis aims to produce gestures synchronized with speech audio, which enables a variety of applications in filmmaking, game design, robotics, and digital human creation~\cite{TALK-Act,showmaker,animateanyone, AudCast,remot}. 
%
%
Considerable enthusiasm~\cite{HA2G_ted, talkshow, diffstyle} is devoted to exploring the elusive relationship between speech audio and holistic body movements at the early stage.

Recently, a few studies~\cite{seeg, co-speech-3, wang2022vpu, beat, EMAGE, sege, SIGGesture} have shifted their focus on the semantic relevance within the gestures and pursue performance improvement by involving speech textual content as an additional condition. \cite{beat, EMAGE} leverage word embedding from text scripts to perform semantic-relevant and rhythm-relevant co-speech motion modeling. The latest studies~\cite{sege, SIGGesture} prefer generative semantic retrieval via large language models (LLM) for better semantic injection.
However, the semantic-gesture correlation studied in these approaches is not as clear as hand gestures with explicit instructions (e.g., numbers and directions), which can be frequently observed in our daily communication.
Verifiably, hand gestures not only improve the efficiency of communication but also facilitate understanding via effective visual patterns~\cite{hands2minds, learn-to-count}, which constructs a crucial component for co-speech generation.
%
%
Therefore, to create life-like human behaviors that can efficiently deliver explicit information, a co-speech semantic gesture generation system is imperative.



However, achieving such a system is not trivial.
\textbf{1)} Most datasets only capture conversational speech-gesture pairs with quite limited frequency of instructional hand gestures (e.g., numbers and directions)
\textbf{2)} The inherent ``many-to-many'' mapping between speech audios and gestures introduces evident complexity to the system, which may hinder consistent and reliable gesture activation and produce undesirable gestures occasionally.
%

To cope with the first challenge, we build a large-scale co-speech human gesture dataset (\textbf{Streamer}), which contains a series of pre-defined hand gestures (e.g., numbers and directions) with accompanying speech audio in Chinese.
%
Concretely, we capture our dataset under a live-streaming scenario in consideration of two points: 
1) Instructional hand gestures are frequently employed in live-streaming scenarios, such as using fingers to indicate quantities or prices.
2) The transcripts for live-streaming are close to our daily life, which is user-friendly for novices.

To address the rest problems, we propose a co-speech \textbf{Gesture} Generation system built upon \textbf{HY}brid modality \textbf{D}iffusion transformer and a cascaded, synchronized \textbf{R}etrieval-\textbf{A}ugmented Generation (\textbf{GestureHYDRA}) which achieves flexible human gesture synthesis and reliable generalization capability under unseen scenarios. 
Our key insight is \textit{to construct a hybrid-modality system for gesture modeling}. 
Specifically, our system takes two kinds of modality signals (i.e., speech audio and gestures) as input, where its training process can be formulated as co-speech gesture generation or motion inpainting when the system receives only a single modality of input.
We identify two advantages of our hybrid-modality system beyond the traditional co-speech gesture generation paradigm:
1) The hybrid training strategy that involves distinct learning tasks is beneficial for the system to improve its gesture modeling ability.
2) Hybrid inputs allow our system to support more flexible editing operations apart from the fundamental co-speech gesture generation, including gesture injection, interpolation, and replacement to fulfill diverse requirements in real-world scenarios.

Based on this system, we first propose Hybrid-modality Diffusion Transformer, a motion-style injective transformer layer to enhance the generalization ability of our system by encoding motion style from dynamic and static dimensions rather than using sparse one-hot embeddings.
Carefully designed masking training strategies are also adopted for effective hybrid training which elegantly integrates different learning tasks via simple masking assignment. 
On the other hand, we also consider the unsuccessful gesture activation issue during the co-speech generation and present a cascaded retrieval-augmented generation strategy to ensure stable gesture activation via individual semantic gesture repository and an injection-timestamp adjustment strategy for adaptive speech-gesture synchronization.

Our contributions are summarized below:
1) We build Streamer, a Chinese semantic gesture dataset containing widely used hand gestures that are compatible with various real-world scenarios.
2) We propose the Hybrid-Modality gesture generation system which is built upon diffusion transformers. It equips the co-speech gesture generation task with versatile functions.
3) We propose a cascaded Retrieval-Augmented Generation scheme for synchronized gesture injection and specific editing that allows automatic gesture injection via adaptive timestamp adjustment.


\section{Related Work}
\subsection{Co-speech Gesture Generation}
Human motion embodies rich and expressive information, rendering human motion analysis a longstanding and extensively studied research problem~\cite{gait,CLaM,EDGE,s3o}.
Co-speech Gesture Generation focuses on understanding the complex relationship between speech audio and holistic body movements. Initially, researchers primarily relied on rule-based methods which required extensive manual effort to model this relationship~\cite{co-speech-rule1, co-speech-rule2, co-speech-rule3, co-speech-rule4,co-speech-rule5}. Data-driven methods~\cite{MDT-A2G,MambaTalk,SGToolkit,diffstyle,DiffSHEG,Audio2Gestures,DiffGesture,co-speech-Rhythmic,co-speech-2,co-speech-4-lstm,co-speech-5,DisCo} replaced rules with diverse models 
trained on large datasets, enabling automatic speech-gesture learning. 
VQ-VAE~\cite{vq-vae,pq-vae,tokengen} and diffusion models~\cite{ddim} have shown strong performance across downstream tasks~\cite{pixel2token,OOSTraj,dmdet} and are now the mainstream choices for co-speech gesture generation.

Some studies~\cite{talkshow,probtalk,diffstyle,DiffSHEG} focus on the rhythmic consistency between speech and generated body movements, particularly gestures, while roughly controlling limited motion styles using a one-shot identity vector.
Recent studies~\cite{sege,SIGGesture,EMAGE} highlighted the importance of semantic gestures in conversational scenes and collected corresponding datasets. 
However, the semantic gestures in these works remain coarse and ambiguous. 
In contrast, our proposed large-scale Streamer dataset focuses on instructional and explicit semantic gestures (e.g., numbers and directions) in real-world scenarios, filling this gap.


\begin{figure*}[t]
\centering
\includegraphics[width=\linewidth]{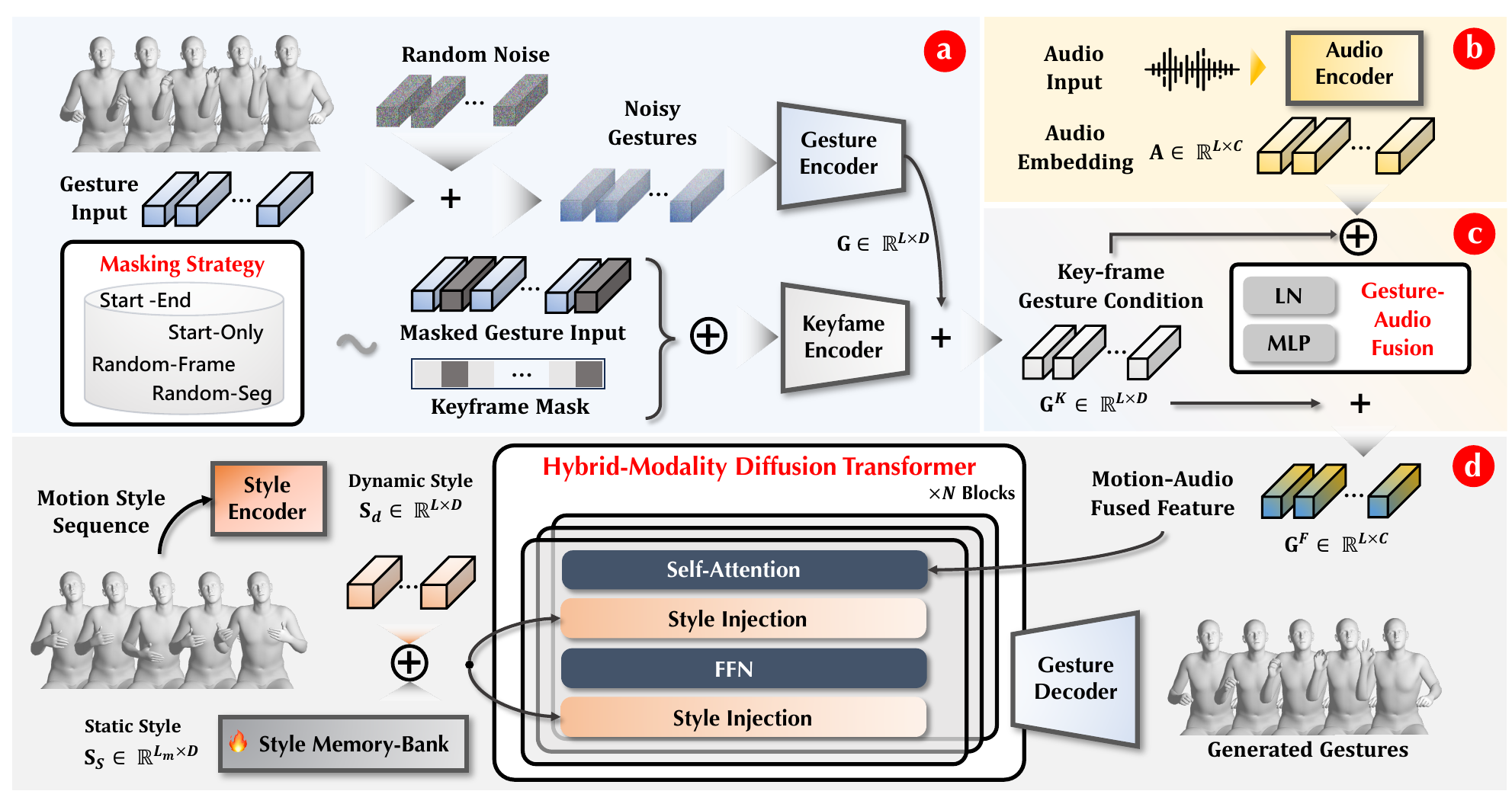}
\caption{
\textbf{Training Framework of GestureHYDRA}. The framework is depicted in four parts: a) \textit{Gesture Encoding}. Noisy gestures are encoded using a gesture encoder. Masking strategies are applied to access key-frame inputs, which are then fed into a key-frame encoder. b) \textit{Audio Encoding}. The speech audio is processed through an audio encoder to extract audio features that guide the gesture generation. c) \textit{Audio-Gesture Fusion}. A fusion module combines the audio features and gesture latents to align both modalities. d) \textit{Gesture Generation}. The fused representation is passed through a transformer layer for gesture generation. A style injective transformer layer is used to preserve the original style of the gestures, ensuring they are both contextually accurate and stylistically consistent.
}
\vspace{-10pt}
\label{fig:pipeline_train}
\end{figure*}


\subsection{Retrieval-Augmented Generation}
Retrieval-Augmented Generation (RAG) is a hybrid approach that enhances generative models with external knowledge from databases. 
By retrieving relevant real-time data, RAG mitigates key limitations of LLMs~\cite{Llama3,Qwen25}, such as knowledge cut-off and hallucination. Recent works~\cite{RAT,mR2AG} also apply RAG to text-to-image/video tasks by incorporating retrieved visual or textual context.

Recently, RAG has also been applied to gesture generation.
ReMoDiffuse~\cite{ReMoDiffuse} explores the utilization of retrieved motions to provide
useful guidance for the denoising process.
For Co-speech generation, Semantic Gesticulator~\cite{sege} merges the retrieved gesture tokens into the initial token sequence via best merging time calculation. 
Similarly, SIGesture~\cite{SIGGesture} inserts the retrieved gestures into the denoised latent space for better generation.
Unlike these methods that directly merge or insert the retrieval results back into the original gesture sequence, our approach relies on a cascaded RAG strategy, which first retrieves a key frame from relevant segments and re-generates expected rhythm and synchronized gestures with the input audio.

\section{Methodology}
The overview of our system is illustrated in Fig.~\ref{fig:pipeline_train}. In this section, we first present the task formulation and the diffusion model preliminaries in Sec.~\ref{sec:task}. 
Then we elaborate on our novelly proposed hybrid modality diffusion transformer (HM-DiT) and its carefully designed training strategies in Sec.~\ref{sec:dit}.
Finally, we introduce the implementation of our semantic-aware retrieval-augmented generation mechanism incorporated within the inference pipeline for reliable semantic gesture activation in Sec.~\ref{sec:rag}.

\subsection{Task Formulation and Preliminaries}
\label{sec:task}
\noindent\textbf{Task Formulation.}
%
Given an audio clip $A = \{a_i\}^{L-1}_{i=0}$, the objective of co-speech gesture generation is to reconstruct the target human gestures $G = \{g_i\}^{L-1}_{i=0}$ including natural body movements and responsive hand gestures with specific semantics, where $L$ denotes the sequence length. 
Concretely, we follow \cite{diffstyle} to utilize a short clip with $S$ frames $G^s = \{g_i\}^{S-1}_{i=0}$ as the initial gestures (a.k.a., seed gestures).

\label{sec:preliminaries}
\noindent\textbf{Preliminaries for Diffusion models.}
Our system is built upon MDM\cite{tevet2022humanmotiondiffusionmodel}, which deploys diffusion transformer blocks for human motion modeling. The diffusion process operates as a Markov chain with $T$ steps, where noisy sequences $g_1, ..., g_T$ originate from real observations $g_0 \sim q(g_0)$. Gaussian noise is gradually applied at each step: $q\left(g_t \mid g_{t-1}\right):=\mathcal{N}\left(g_t ; \sqrt{1-\beta_t} g_{t-1}, \beta_t \mathbf{I}\right)$. 
%
%
The reverse process aimed to reconstruct the original gestures from distorted gesture sequences. Unlike the image generation task~\cite{ldm}, MDM prefers to predict the original sequence $g_0$ instead of the noise by using a learnable deep learning network, denoted by $\overline{g}_0=S_\theta\left(g_t, t, c\right)$, where $c$ denotes the conditions.
The training goal of this model is designed to minimize the mean squared error (MSE) between the original sequence $g_0$ and the predicted result $\overline{g}_0$, expressed as $L_t=\mathbb{E}_{g_0, \epsilon, t}\left[g_0-S_\theta\left(g_t, t, \mathrm{c}\right)\right]$.

\begin{figure}[t]
\centering
\includegraphics[width=\linewidth]{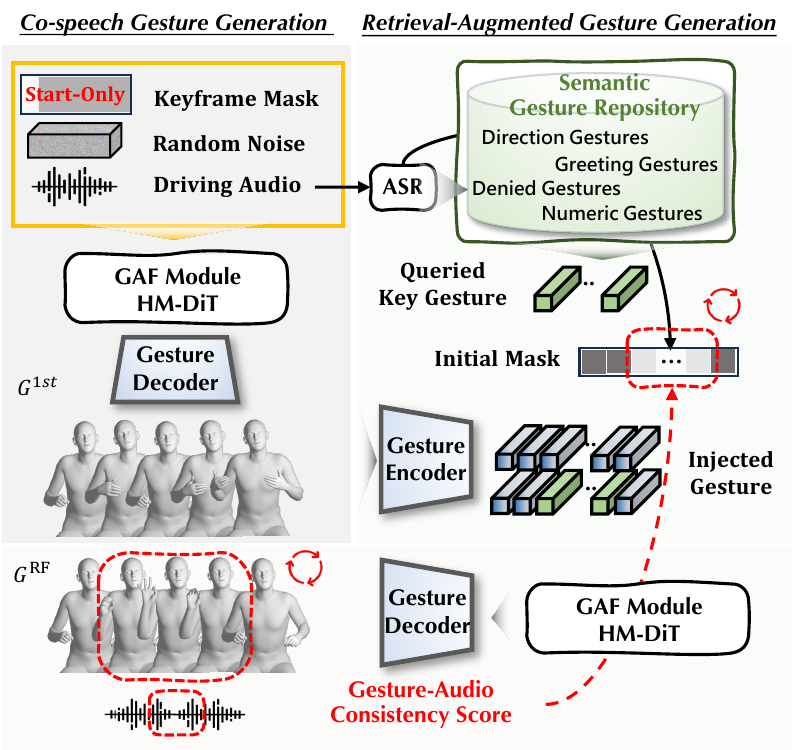}
\caption{
\textbf{Cascaded-Synchronized Retrieval-Augmented Generation}. By using the RAG scheme, we can refine the co-speech results via relevant gesture retrieval and automatic synchronization by progressively updating the injection timestamp.
}
\label{fig:pipeline_infer}
\end{figure}

\subsection{Hybrid-Modality Diffusion Transformer}
\label{sec:dit}
Unlike the one-to-many correspondence observed between speech audio and lip movements, the relationship between audio and gestures is characterized by a many-to-many mapping, where a single gesture can be associated with diverse acoustic patterns.
This inherent mapping significantly holds back the model from learning whether to activate the gesture, leading to occasional activation failures or unnecessary activation from similar words. To construct a more robust and generalized system, we introduce our hybrid-modality diffusion transformer (HM-DiT) backbone and motion-style injection module.

\noindent\textbf{Masking Strategies For Hybrid Training.} Considering that unexpected gesture activation may interrupt human gesture creation, our HM-DiT is designed to take two modalities of signals (i.e., audio and human gestures) as input.
Moreover, we carefully design four masking strategies for human gestures $G$ as depicted in Fig.~\ref{fig:pipeline_train}(a) to simulate real-world scenarios that require gesture editing and each strategy appears with equal probability.
Among them, ``Start-Only'' denotes masking all frames except the seed gestures $G^s$, which is identical to the standard co-speech gesture generation setting. 
Similarly, ``Start-End'' denotes an additive symmetry condition at the end of $G$ and follows the speech-driven motion in-betweening procedure.
In terms of ``Random-Frame'' and ``Random-Seg'', we focus on the capability of global modeling and consecutive segment synthesis. Notably, the training procedure under these two masking strategies will switch into unconditional motion in-betweening when the speech guidance scale in the classifier-free guidance setup turns to zero.

\noindent\textbf{Hybrid Gesture Synthesis Training Procedure.} During training, we first produce noisy gestures $G'$ perturbed by Gaussian noise $\epsilon$ and leverage a Gesture Encoder $\bm{E}_G$ to extract noisy gesture latent $\mathbf{G} \in \mathbb{R}^{L \times D}$.
Once the masking strategy is specified, we derive the key-frame gesture $G^m$ from $G$ by using the training mask $M = \{m_i\}^{L-1}_{i=0}, m_i \in \{0, 1\}$ and then perform channel-wise concatenation on $G^m$ and $M$. Then a KeyFrame Encoder $\bm{E}_K$ similar to $\bm{E}_G$ is employed to extract key-frame gesture latent, which is subsequently added to the noisy gesture latent $\mathbf{G}$ to form a fused latent $\mathbf{G^{K}} \in \mathbb{R}^{L \times D}$.
In the next step, we turn our attention to the speech audio aspect.
We first leverage a widely used Audio Encoder~\cite{wav2vec} $\bm{E}_A$ to extract audio embeddings $\mathbf{A} \in \mathbb{R}^{L \times C}$.
Then we follow \cite{DiffSHEG} to perform cross-modality alignment by using a simple gesture-audio fusion (GAF) module with a residual block instead of cross-attention layers~\cite{attention}, which consists of a layer normalization (LN) operation and a multiple-layer perception (MLP). The fused latent $\mathbf{G}^{F} \in \mathbb{R}^{L \times C}$ can be defined as:

\begin{equation}
    \mathbf{G^{F}} = \mathbf{G^{K}} + \text{GAF}(\mathbf{A} \oplus \mathbf{G^{K}}),
\end{equation}
where $\oplus$ is the channel-wise concatenation operation. The fused cross-modality latent $\mathbf{G}^{F}$ is then fed into the transformer layer for gesture prediction.

\noindent\textbf{Motion-Style Injective Transformer Layer.} We initialize our transformer layer following~\cite{attention} and observe that generated gestures fail to preserve their original styles without further fine-tuning on individual data.
Most previous approaches~\cite{diffstyle, probtalk} adopt one-hot embeddings for speaker identity guidance, however, they cannot provide generalized performance, especially when the number of identities keeps increasing.

Recently, a zero-shot talking face generation approach~\cite{resyncer} built upon speech-driven facial mesh animation proposes to improve its generalization ability by using an additional motion sequence as talking style reference. We take inspiration from it and propose to involve an additional gesture sequence and newly designed injective transformer layers for a more robust motion-style injection. 
Specifically, the motion-style injective transformer layer shown in Fig.~\ref{fig:pipeline_train}(d) adopts a standard combination of a self-attention layer and a feed-forward network (FFN) layer with two style injection layers placed right after them, respectively. The style injection layer is designed to integrate the benefits of both dynamic motion components and static motion components, where the dynamic component denotes the motion-style embedding $\mathbf{S_d} \in \mathbb{R}^{L \times D}$ derived from a style encoder $\bm{E}_S$ representing captured motion-style information from external input, while the static component is an internal learnable motion memory-bank $\mathbf{S_s} \in \mathbb{R}^{L_m \times D}$ similar as ~\cite{vpgc, vprq}, memorizing all the motion-styles within the training data.
The final motion-style embedding $\mathbf{S} \in \mathbb{R}^{(L_m + L) \times D}$ is achieved by concatenating $\mathbf{S_d}$ and $\mathbf{S_s}$ along the temporal dimension and the style-injection attention can be formulated as:
\begin{equation}
    \text{Att}_{style} = \text{softmax}\left(\frac{\mathbf{G^{F'}} {\mathbf{S}}^\top}{\sqrt{c}}\right) {\mathbf{S}},
\end{equation}
where $\mathbf{G^{F'}}$ denotes the noisy gesture latent outputs from the self-attention layer or the FFN layer.
Notably, for each speaker identity, we carefully select a gesture sequence distinct from $G$ as the motion style sequence during training to mitigate gesture leakage. Furthermore, a segment shuffle operation is conducted to prevent excessive dependency on the motion-style patterns captured from the entire sequence.

\begin{table*}[t]
    \centering
    \small  
    \caption{\textbf{Quantitative evaluation on the Streamer Datasets}.
    The $\downarrow$ means the lower, the better. 
    The $\uparrow$ means the higher, the better.}
    \label{tab:main_result}

    \newcolumntype{Y}{>{\raggedright\arraybackslash}X} 
    \newcolumntype{Z}{>{\centering\arraybackslash}X}  
    \newcolumntype{M}[1]{>{\centering\arraybackslash}m{#1}}  

    \begin{tabularx}{\linewidth}{lM{3cm}ZZZZZZ} 
        \toprule
        Dataset & System & FGD $\downarrow$ & $\Delta BC$ $\downarrow$ &
        SAR $\uparrow$  &
        SMD-L1  $\downarrow$  & SMD-DTW  $\downarrow$   \\ 
        \midrule
        \multirow{5}{*}{Seen Identity} 
        & GT               & - & 0 (0.697) & - & - & - \\
        \cmidrule(lr){2-7}
        & TalkSHOW         & 51.50 & 0.062 (0.759)  & 61.49\% & 0.161 & 32.11 \\
        & Probtalk         & 50.33 & 0.007 (0.690)  & 72.29\% & 0.120 & 22.37 \\
        & DSG              & 54.59 & 0.072 (0.769)  & 73.03\% & 0.116 & 22.61 \\
        & Ours             & \textbf{3.24}  & \textbf{0.003} (0.700)  & \textbf{84.82\%} & \textbf{0.107} & \textbf{20.70} \\
        
        \midrule
        \multirow{5}{*}{Unseen Identity} 
        & GT               & - & 0 (0.681) & - & - & - \\
        \cmidrule(lr){2-7}
        & TalkSHOW         & 75.35 & 0.085 (0.766) & 31.81\% & 0.210 & 41.00 \\
        & Probtalk         & 63.74 & 0.030 (0.711) & 66.08\% & 0.174 & 33.26 \\
        & DSG              & 61.94 & 0.091 (0.772) & 68.77\% & 0.160 & 30.77 \\
        & Ours             & \textbf{15.43} & \textbf{0.027} (0.708) & \textbf{81.36\%} & \textbf{0.143} & \textbf{27.73} \\
        \bottomrule
    \end{tabularx}
\end{table*}

\noindent\textbf{Training Objectives.}
In addition to MSE loss $\mathcal{L}_t$ described in \ref{sec:preliminaries}, we also incorporate the velocity loss~\cite{DiffSHEG} $L_{vec}$ and a 3D key point loss $L_{kp}$ to ensure similar human motion prediction to stable and accurate 3D positions, which provides effective guidance for downstream applications such as controllable video generation. 
Since we directly predict the original gestures $\overline{x}_0$ rather than the noise, it is convenient to calculate these auxiliary losses. 
%
%
The velocity loss and the 3D keypoint loss are both formulated using L1 distance. The velocity loss measures discrepancies in velocities, while the 3D keypoint loss quantifies the gap between predicted and ground-truth 3D keypoints.
The total loss function is a weighted sum of the three losses:
\begin{equation}
\mathcal{L}=\lambda_t \mathcal{L}_t+\lambda_{vec} \mathcal{L}_{vec}+\lambda_{kp} \mathcal{L}_{kp},
\end{equation} where $\lambda_t=10$, $\lambda_{vec}=1$ and $\lambda_{kp}=1$.

\begin{table}[t]
    \centering
    \caption{\textbf{Quantitative evaluation on the SHOW Dataset}. 
    The $\downarrow$ means the lower, the better.}
    \label{tab:show}

    \newcolumntype{Y}{>{\raggedleft\arraybackslash}X}
    \newcolumntype{Z}{>{\centering\arraybackslash}X}
    \begin{tabularx}{\linewidth}{lYY} 
        \toprule
        System   & FGD $\downarrow$ & $\Delta BC$ $\downarrow$ \\ 
        \midrule
        GT       & -                & 0 (0.847)       \\ 
        \cmidrule(lr){1-3}
        TalkSHOW & 6.04             & 0.023 (0.870)   \\ 
        Probtalk & 5.46             & 0.046 (0.801)   \\ 
        DSG      & 6.33             & 0.022 (0.869)   \\ 
        Ours     & \textbf{3.68}    & \textbf{0.006} (0.853) \\ 
        \bottomrule
    \end{tabularx}
\end{table}

\subsection{Cascaded-Synchronized RAG}
\label{sec:rag}
At the inference stage, our system supports two gesture prediction modes as shown in Fig.~\ref{fig:pipeline_infer}: 1) co-speech gesture generation based on the ``Start-Only'' masking strategy. 2) retrieval-augmented generation (RAG) with a specific key-frame mask. We provide a detailed description below to introduce these two modes.

\noindent\textbf{Semantic Gesture Repository.} Given the captured dataset with semantic hand gestures, we manually build a semantic gesture repository for every single identity according to the 18 pre-defined gestures. 
Each semantic gesture category comprises at least one clip lasting about one second, which serves as a gesture prototype. 
Moreover, we carefully annotate a key-frame with an ideal gesture from each clip for flexible gesture activation. 
Notably, we also provide an automatic annotation approach for any in-the-wild data. 
Please refer to the supplementary file for more details.

\noindent\textbf{Adaptive Key Gesture Injection.} Imagine that we perform a co-speech generation with driving audio and its output $G^{1st}$ may suffer from an unsuccessful semantic gesture activation, leading to the degradation of naturalness and realism.
Thanks to the capability built by our HM-DiT, we can intuitively switch to the RAG generation mode with hybrid-modality inputs for gesture refinement. 
Here we first leverage automatic speech recognition (ASR) techniques to identify the gesture-relevant phrase as well as its corresponding segment $G^{1st}_{t_s:t_e}$ that requires refinement and the phrase is subsequently used to retrieve the corresponding gesture clip $G^{sem}$ from the semantic gesture repository. However, how to identify the injection timestamp and how to activate the expected gestures have not been well investigated yet.

A recent study~\cite{sege} attempts to obtain injection timestamps by detecting audio beats corresponding to the stroke phase of gestures and merges the retrieved gestures by generating linking segments with rhythmic gestures.
Different from \cite{sege}, we observe that the gesture clip $G^{sem}$ retrieved from the semantic gesture repository is likely to have a different ongoing beat with the speech audio, especially in-the-wild ones. Thus, we prefer to inject \textit{the annotated ideal gesture frame} rather than the entire segment, which enforces the generated rhythm to rely more on the actual input audio instead of $G^{sem}$.
To identify a suitable injection timestamp for the ideal gesture frame, we propose an adaptive timestamp adjustment strategy based on the audio-gesture consistency score~\cite{liu2022learning, learndance} to guarantee our generated gesture process is synchronized with the semantic phrase, exempt from delay or rush.
Specifically, we set the midpoint of $t_s$ and $t_e$ as the initial timestamp and repeatedly perform the RAG generation process while recording a temporary best score (see Section 4.5). 
We keep moving the timestamp following the binary search procedure until we find the one with the best score.
According to the final best score, we achieve the most appropriate injection timestamp and a satisfactory generated result.

\section{Experiments}
\subsection{Datasets}

In this section, we conducted comprehensive experiments on the SHOW~\cite{talkshow} dataset and the proposed Streamer dataset to validate the effectiveness of our method.

\noindent\textbf{SHOW Dataset}.The SHOW dataset comprises 27 hours of talking videos from 4 actors, which includes reconstructed SMPL-X parameters of the whole body and synchronized speech data. 
The dataset emphasizes the synchronization between speech rhythm and gestures while lacking explicit semantic information. 
Following~\cite{probtalk}, we selected 10,706 clips as the training set and 1,313 clips as the test set.

\noindent\textbf{Streamer Dataset}. We recorded monocular videos of a solo anchor in a well-lit studio environment using a high-definition camera. 
Given text with semantic information, the anchor reads in Mandarin at a constant pace. 
During moments without semantics, the anchor naturally and realistically performs gestures according to his style. 
When semantics are present, the anchor needs to make gestures that are both natural and semantically aligned with the speech.
Specifically, the Streamer dataset contains 281 actors with a total of 20,969 ten-second clips. 
To evaluate the generalization capability of our proposed model, we first selected data from 10 individuals in the dataset, totaling 998 clips, as the unseen identity test set. 
Then, we randomly selected 920 clips from the remaining data as the test set for seen identities. 
Finally, the rest of the data was used as the training set.
For more detailed information about the dataset, please refer to our supplementary materials.


\subsection{Experimental Settings}
\noindent\textbf{Implementation Details.} Our model is trained on four 40G A100 GPUs for about 3 days. The training process starts with a 120k-step pre-training using a batch size of 512, excluding the 3D keypoint loss, and continues with an additional 30k steps while incorporating the 3D keypoint loss. 
%
%
Due to the slow forward speed of the SMPL-X model, we apply the 3D keypoint loss only on sparsely selected frames with a ratio of 1/8 at most.
The models are trained with 1000 noising steps and a cosine noise schedule. During inference, we use a DDIM sampler with 50 denoising steps.

\begin{figure*}[t]
  \centering
  \includegraphics[width=0.97\linewidth]{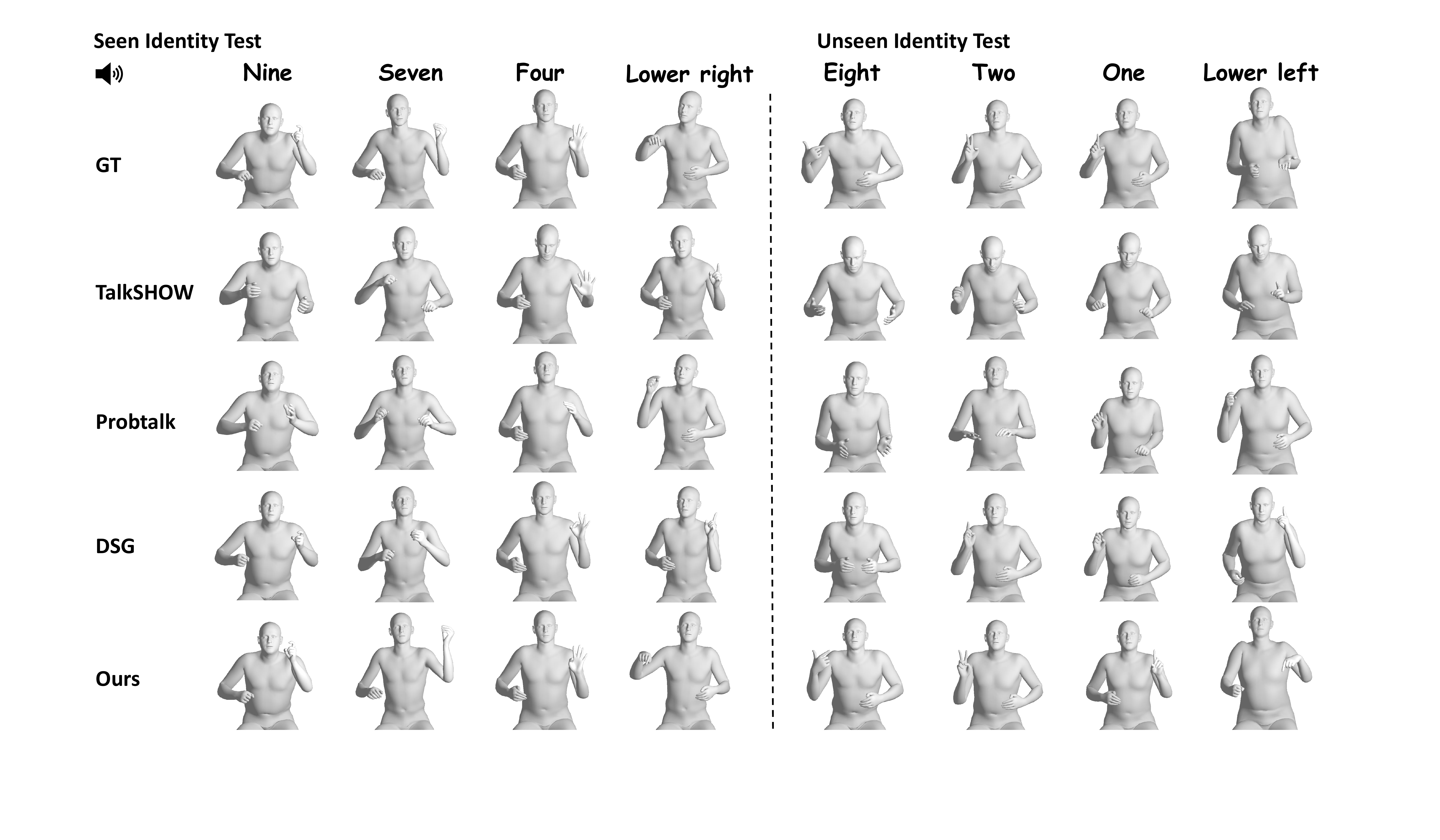}
  \caption{\textbf{Qualitative results on the Streamer dataset}. 
Compared to other SOTAs, our approach achieves accurate semantic gesture synthesis in both settings.}
  \label{fig:main_result} 
\end{figure*}

\noindent\textbf{Comparing Baselines.} We select several state-of-the-art approaches, including TalkSHOW~\cite{talkshow}, Probtalk~\cite{probtalk}, and DiffuseStyleGesture (DSG)~\cite{diffstyle} as comparison methods. 
Both TalkShow and ProbTalk follow a VQ-VAE~\cite{vq-vae} paradigm, and TalkShow uses an autoregressive approach to predict motion distributions, while ProbTalk adopts a MaskGIT-like generative paradigm.
DSG, on the other hand, is a diffusion-based approach.
We use their pre-trained models for evaluation on the SHOW dataset and conduct training on the Streamer dataset using their released code.
To ensure fairness, all comparisons are limited to the body and hands, excluding the face expressions.

\noindent\textbf{Evaluation Metrics.} We evaluate the performance of various methods from multiple perspectives. First, following~\cite{probtalk}, we select two commonly used metrics: Fr\'{e}chet Gesture Distance (FGD)~\cite{co-speech-1} and Beat Consistency Score (BC)~\cite{learndance}. Notably, we calculate the absolute difference for BC scores between predicted gestures and ground-truth (denoted as $\Delta BC$) for convenient comparison .
%
%
Recent studies~\cite{SIGGesture,DiffSHEG} have pointed out that these metrics cannot fully and accurately reflect the quality of generated motions.
Therefore, we propose two new evaluation metrics: Semantic Activation Rate (SAR) and Semantic Motion Distance (SMD) to assess the generation quality of semantic gestures from the Streamer dataset.
SAR provides the success rate of generating correct gestures within the annotated semantic intervals, based on manual annotations.
In terms of SMD, we first obtain the corresponding 3D key points from the generated gestures and the ground-truth gestures. 
Then we calculate the L1 distance and Dynamic Time Warping (DTW) distance on the upper limbs within each semantic segment, where DTW can handle motion sequences with varying rhythms.
We randomly select 450 clips from each of the unseen and seen test sets to evaluate the SAR and SMD metrics.
Notably, generated samples that are correctly activated but performed with the hand opposite to the ground truths are filtered out when calculating SMD.

\subsection{Evaluation Results}
In this section, our results do not incorporate our proposed cascaded-synchronized RAG for result enhancement.

\noindent\textbf{Quantitative Results.}
The quantitative comparison results between our system and SOTA methods are shown in Table~\ref{tab:main_result} and Table~\ref{tab:show}. 
Since the SHOW dataset lacks explicit semantic content, we only compare the FGD and $\Delta BC$ metrics on the SHOW dataset.
It can be observed that our method surpasses existing SOTA methods on both the FGD and $\Delta BC$ metrics, highlighting the superiority of the gestures generated by our approach in terms of feature distribution and beat matching.


On the Streamer test dataset, in addition to reporting the FGD and $\Delta BC$ metrics, we also provide the SAR, SMD-L1, and SMD-DTW metrics to evaluate the quality of semantic gesture generation.
On the seen test set, our method achieves the best results on the FGD metric, thanks to the well-designed hybrid learning process. 
For the $\Delta BC$ score metric, the generated motions from TalkSHOW and DSG suffer from noticeable jitter, with such anomalies and motion artifacts leading to higher $\Delta BC$ values. 
%
Compared to other SOTAs, our method achieves the best $\Delta BC$ score, indicating that our generated gesture sequences are well synchronized with the input audio.
As for the quality of semantic gesture generation, our system benefits from a well-designed masking strategy, enabling it to effectively capture the relationships between different types of motion sequences while simultaneously modeling the correlation between speech and motion. 
As a result, our system achieves the best performance in both SAR and SMD.

On the unseen test set, our method also outperforms existing SOTA methods on all metrics, demonstrating its superior generalization capability. 
In contrast, the VQ-VAE-based counterparts suffer from limited generalization due to their limited codebooks. 
In summary, our system can generate motions that reach a realistic level, with robust performance in rhythm alignment and semantic gesture generation quality on both seen and unseen data.

\begin{figure}[t]
    \centering
    \includegraphics[width=1\linewidth]{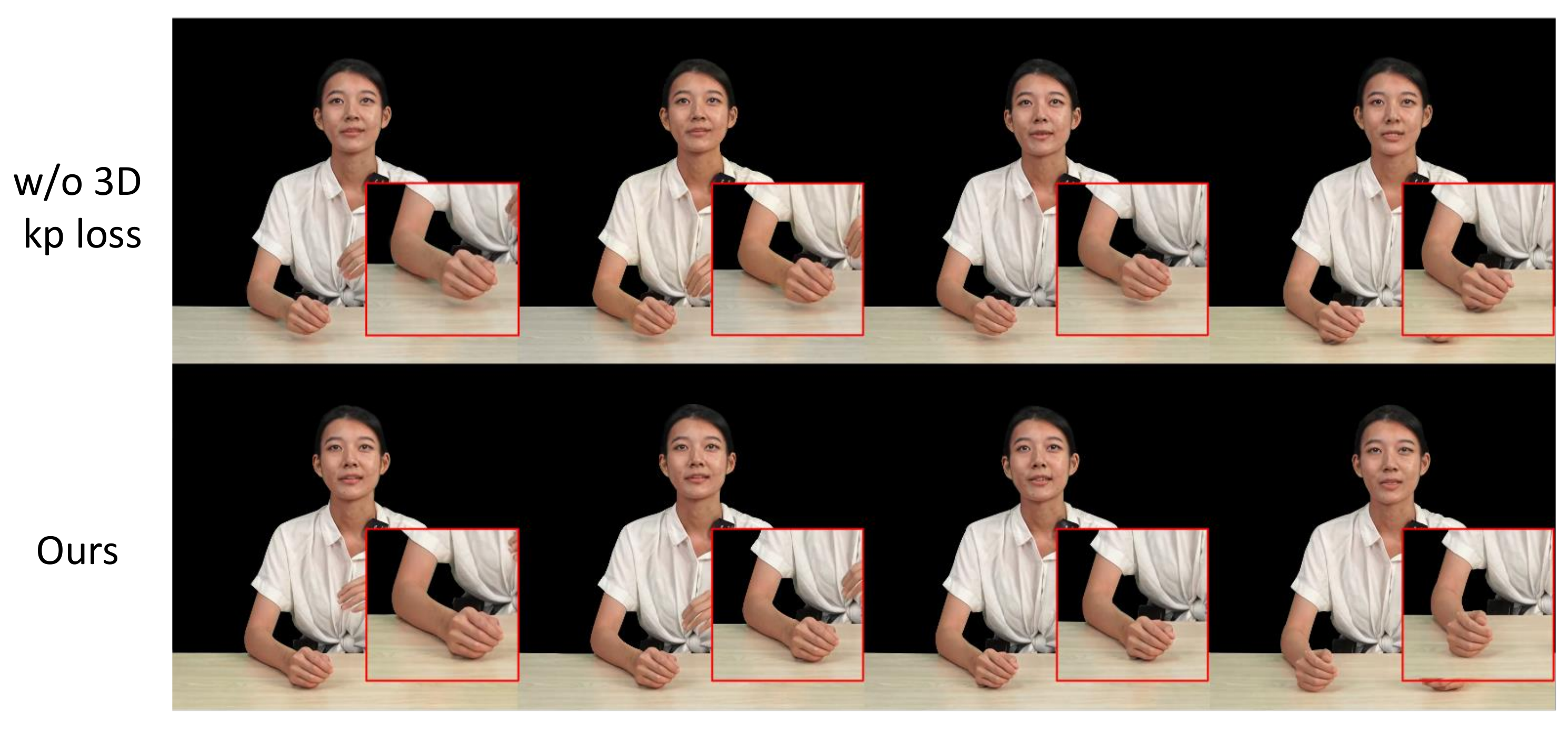}
    \caption{\textbf{Comparisons of Video Generation Results.} \textit{Top}: w/o 3D keypoint loss. \textit{Bottom}: with 3D keypoint loss. Adding the 3D keypoint loss improves the stability of hand-to-table interactions.} 
    \label{fig:enter-label}
\end{figure}

\noindent\textbf{Qualitative Results.}
We strongly recommend watching the provided demo video (including the visualization results of the SHOW dataset) for a more intuitive experience. 
Fig.~\ref{fig:main_result} shows the generated results on the Streamer dataset, including results from both the seen and unseen identity test sets.
On the seen identity test set, our method successfully generates correct semantic gestures in response to the speech input, whereas other methods either fail to trigger semantic gestures (see column 2), produce incorrect ones (see column 3), or yield gestures lacking sufficient accuracy (see row 4, column 1). 
On the more challenging unseen identity setting, our method consistently manages to activate correct semantic gestures, while other methods still struggle to produce precise semantic gestures. 
%

Notably, human evaluation is also provided, please refer to our supplementary materials for more details.

\subsection{Ablation study}
To evaluate the contributions of crucial components in our system, we conduct an ablation study on the Streamer unseen test set and the results are reported in Table~\ref{table:abla}.
Specifically, we construct three variants by removing the corresponding component from our full model, denoted as ``w/o mask strategy'', ``w/o motion style'' and ``w/o 3D kp loss''. Please note that ``w/o mask strategy'' denotes that our mask strategy merely supports ``Start-Only'' and the pipeline is converted into pure co-speech generation.
The results reveal that the removal of the mask strategy, motion style, and 3D keypoint loss, causes a significant performance drop on the 3D keypoint distance between the generated semantic gestures and the GTs, which indicates that all three designs contribute to generating accurate semantic gestures.

To clearly demonstrate the effectiveness of ``3D kp loss'', we additionally perform a study on a downstream task (i.e., human video generation), which produces corresponding human videos based on our generated co-speech gestures by using a human video generation model~\cite{animateanyone}.
As shown in Fig.~\ref{fig:enter-label}, our full model is able to synthesize a stable human body with satisfying hand-to-table interactions instead of floating elbows as in the results of ``w/o 3D kp loss'',
which further facilitates the effectiveness of our $L_{kp}$.

\subsection{Analysis of Adaptive Key Gesture Injection}
As aforementioned, we calculate $\Delta BC$ score between the generated semantic part and the gesture prototype in the database as the temporary best score.
However, selecting the fixed insertion timestep based on ASR is also an alternative approach. 
To validate the advantages of our proposed  adaptive injection strategy, 
we conducted comparative experiments on several samples where activation failed. 
The variants include: without insertion strategy (w/o Injection), using the intermediate position from ASR as fixed insertion points (Vanilla Injection), and our proposed Adaptive Key Gesture Injection (Adaptive Injection). The experimental results reported in Table~\ref{table:inject} demonstrate that the predicted semantic gestures exhibit better alignment with the speech by utilizing the $\Delta BC$ score as a metric for synchronization, which further enhances the performance of semantic gesture generation.

\begin{table}[t]
\centering
\caption{The results of the ablation study}
\label{tab:ablation}
\resizebox{0.48\textwidth}{!}{ 
\begin{tabular}{lccc}
\hline
System   & FGD $\downarrow$ & SMD-L1 $\downarrow$  & SMD-DTW $\downarrow$  \\ \hline
w/o mask strategy & 15.76     & 0.156           & 29.71   \\
w/o motion style  & 20.31     & 0.155           & 29.51   \\
w/o 3D kp loss    & \textbf{14.80}  & 0.154     & 29.68   \\
Ours              & 15.43  & \textbf{0.143}  & \textbf{27.73}           \\ \hline
\end{tabular}
}
\label{table:abla}
\end{table}

\begin{table}[t]
\centering
\caption{The analysis of Adaptive Key Gesture Injection}
\label{tab:ablation}
\resizebox{0.45\textwidth}{!}{ 
\small 
\begin{tabular}{lcc}
\hline
variants  & SMD-L1 $\downarrow$  & SMD-DTW $\downarrow$  \\ \hline
w/o Injection       & 0.176      & 30.46   \\
Vanilla Injection   & 0.155      & 27.45   \\
Adaptive Injection  & \textbf{0.138}  & \textbf{26.88}  \\ \hline
\end{tabular}
}
\label{table:inject}
\end{table}





\section{Conclusion}
In this work, we propose GestureHYDRA, a co-speech gesture generation system built upon hybrid-modality diffusion transformers.
We highlight several excellent properties of our system:
1) Our system receives hybrid-modality inputs, which allows hybrid learning tasks for model enhancement and a flexible gesture processing scheme.
2) Our proposed motion-style injective transformer layer represents the motion style with dynamic and static aspects, which achieves robust performance under unseen scenarios.
3) Our novelly designed cascaded-synchronized retrieval-augmented generation enables automatic key gesture injection via adaptive timestamp adjustment.


\section*{Acknowledgments}
This work is supported by the National Nature Science Foundation of China (62425114, 62121002, U23B2028, 62232006, 62472395).
We acknowledge the support of the GPU cluster built by the MCC Lab of Information Science and Technology Institution, USTC. 
We also thank the USTC Supercomputing Center for providing computational resources for this project.

{
    \small
    \bibliographystyle{ieeenat_fullname}
    \bibliography{sample-base}

@String(CVPR= {IEEE Conf. Comput. Vis. Pattern Recog.})

@String(ICCV= {Int. Conf. Comput. Vis.})

@String(ECCV= {Eur. Conf. Comput. Vis.})

@String(NIPS= {Adv. Neural Inform. Process. Syst.})

@String(ICLR = {Int. Conf. Learn. Represent.})

@String(IJCAI = {IJCAI})

@String(AAAI = {AAAI})

@String(CVPR  = {CVPR})

@String(ICCV  = {ICCV})

@String(ECCV  = {ECCV})

@String(NIPS  = {NeurIPS})

@String(ICLR  = {ICLR})

@String{Computer = "{IEEE} Computer" }

@String{Springer = "Springer-Verlag" }

@inproceedings{remot,
  author       = {Quanwei Yang and
                  Xinchen Liu and
                  Wu Liu and
                  Hongtao Xie and
                  Xiaoyan Gu and
                  Lingyun Yu and
                  Yongdong Zhang},
  title        = {{REMOT:} {A} Region-to-Whole Framework for Realistic Human Motion
                  Transfer},
  booktitle    = {{ACM} Multimedia},
  pages        = {1128--1137},
  publisher    = {{ACM}},
  year         = {2022}
}

@inproceedings{EDGE,
  author       = {Jonathan Tseng and
                  Rodrigo Castellon and
                  C. Karen Liu},
  title        = {{EDGE:} Editable Dance Generation From Music},
  booktitle    = {{CVPR}},
  pages        = {448--458},
  publisher    = {{IEEE}},
  year         = {2023}
}

@inproceedings{s3o,
  title={S3O: a dual-phase approach for reconstructing dynamic shape and skeleton of articulated objects from single monocular video},
  author={Zhang, Hao and Li, Fang and Rawlekar, Samyak and Ahuja, Narendra},
  booktitle={Proceedings of the 41st International Conference on Machine Learning},
  pages={59191--59209},
  year={2024}
}

@article{pixel2token,
  title={From pixels to tokens: Byte-pair encoding on quantized visual modalities},
  author={Zhang, Wanpeng and Xie, Zilong and Feng, Yicheng and Li, Yijiang and Xing, Xingrun and Zheng, Sipeng and Lu, Zongqing},
  journal={arXiv preprint arXiv:2410.02155},
  year={2024}
}

@inproceedings{OOSTraj,
  title={OOSTraj: Out-of-Sight Trajectory Prediction With Vision-Positioning Denoising},
  author={Zhang, Haichao and Xu, Yi and Lu, Hongsheng and Shimizu, Takayuki and Fu, Yun},
  booktitle={Proceedings of the IEEE/CVF Conference on Computer Vision and Pattern Recognition},
  pages={14802--14811},
  year={2024}
}

@inproceedings{dmdet,
  author       = {Shoufa Chen and
                  Peize Sun and
                  Yibing Song and
                  Ping Luo},
  title        = {DiffusionDet: Diffusion Model for Object Detection},
  booktitle    = {{ICCV}},
  pages        = {19773--19786},
  publisher    = {{IEEE}},
  year         = {2023}
}

@inproceedings{pq-vae,
  author       = {Hanti Wu and
                  Markus Flierl},
  title        = {Learning Product Codebooks Using Vector-Quantized Autoencoders for
                  Image Retrieval},
  booktitle    = {GlobalSIP},
  pages        = {1--5},
  publisher    = {{IEEE}},
  year         = {2019}
}

@article{tokengen,
  title={Visual autoregressive modeling: Scalable image generation via next-scale prediction},
  author={Tian, Keyu and Jiang, Yi and Yuan, Zehuan and Peng, Bingyue and Wang, Liwei},
  journal={Advances in neural information processing systems},
  volume={37},
  pages={84839--84865},
  year={2024}
}

@inproceedings{CLaM,
  author       = {Xiaodong Chen and
                  Kunlang He and
                  Wu Liu and
                  Xinchen Liu and
                  Zheng{-}Jun Zha and
                  Tao Mei},
  title        = {CLaM: An Open-Source Library for Performance Evaluation of Text-driven
                  Human Motion Generation},
  booktitle    = {{ACM} Multimedia},
  pages        = {11194--11197},
  publisher    = {{ACM}},
  year         = {2024}
}

@inproceedings{gait,
  author       = {Jinkai Zheng and
                  Xinchen Liu and
                  Wu Liu and
                  Lingxiao He and
                  Chenggang Yan and
                  Tao Mei},
  title        = {Gait Recognition in the Wild with Dense 3D Representations and {A}
                  Benchmark},
  booktitle    = {{CVPR}},
  pages        = {20196--20205},
  publisher    = {{IEEE}},
  year         = {2022}
}

@inproceedings{AudCast,
  title = {AudCast: Audio-Driven Human Video Generation by Cascaded Diffusion Transformers},
  author = {Guan, Jiazhi and  Wang, Kaisiyuan and Xu, Zhiliang and Yang, Quanwei and Sun, Yasheng and He, Shengyi and Liang, Borong and Cao, Yukang and Li, Yingying and Feng, Haocheng and Ding, Errui and Wang, Jingdong and Zhao, Youjian and Zhou, Hang and Liu, Ziwei},
  booktitle = {Proceedings of the IEEE/CVF Conference on Computer Vision and Pattern Recognition (CVPR)},
  year = {2025}
}

@inproceedings{TALK-Act,
  author       = {Jiazhi Guan and
                  Quanwei Yang and
                  Kaisiyuan Wang and
                  Hang Zhou and
                  Shengyi He and
                  Zhiliang Xu and
                  Haocheng Feng and
                  Errui Ding and
                  Jingdong Wang and
                  Hongtao Xie and
                  Youjian Zhao and
                  Ziwei Liu},
  title        = {TALK-Act: Enhance Textural-Awareness for 2D Speaking Avatar Reenactment
                  with Diffusion Model},
  booktitle    = {{SIGGRAPH} Asia},
  pages        = {109:1--109:11},
  publisher    = {{ACM}},
  year         = {2024}
}

@inproceedings{showmaker,
  author       = {Quanwei Yang and
                  Jiazhi Guan and
                  Kaisiyuan Wang and
                  Lingyun Yu and
                  Wenqing Chu and
                  Hang Zhou and
                  ZhiQiang Feng and
                  Haocheng Feng and
                  Errui Ding and
                  Jingdong Wang and
                  Hongtao Xie},
  title        = {ShowMaker: Creating High-Fidelity 2D Human Video via Fine-Grained
                  Diffusion Modeling},
  booktitle    = {NeurIPS},
  year         = {2024}
}

@article{Llama3,
  author       = {Abhimanyu Dubey and
                  Abhinav Jauhri
                  et al.},
  title        = {The Llama 3 Herd of Models},
  journal      = {CoRR},
  volume       = {abs/2407.21783},
  year         = {2024}
}

@article{Qwen25,
  author       = {An Yang and
                  Baosong Yang and
                  Beichen Zhang et al.},
  title        = {Qwen2.5 Technical Report},
  journal      = {CoRR},
  volume       = {abs/2412.15115},
  year         = {2024}
}

@article{RAT,
  author       = {Zhennan Chen and
                  Yajie Li and
                  Haofan Wang and
                  Zhibo Chen and
                  Zhengkai Jiang and
                  Jun Li and
                  Qian Wang and
                  Jian Yang and
                  Ying Tai},
  title        = {Region-Aware Text-to-Image Generation via Hard Binding and Soft Refinement},
  journal      = {CoRR},
  volume       = {abs/2411.06558},
  year         = {2024}
}

@article{mR2AG,
  author       = {Tao Zhang and
                  Ziqi Zhang and
                  Zongyang Ma and
                  Yuxin Chen and
                  Zhongang Qi and
                  Chunfeng Yuan and
                  Bing Li and
                  Junfu Pu and
                  Yuxuan Zhao and
                  Zehua Xie and
                  Jin Ma and
                  Ying Shan and
                  Weiming Hu},
  title        = {mR\({}^{\mbox{2}}\)AG: Multimodal Retrieval-Reflection-Augmented Generation
                  for Knowledge-Based {VQA}},
  journal      = {CoRR},
  volume       = {abs/2411.15041},
  year         = {2024}
}

@inproceedings{vq-vae,
  author       = {A{\"{a}}ron van den Oord and
                  Oriol Vinyals and
                  Koray Kavukcuoglu},
  title        = {Neural Discrete Representation Learning},
  booktitle    = {{NIPS}},
  pages        = {6306--6315},
  year         = {2017}
}

@inproceedings{ddim,
  author       = {Jiaming Song and
                  Chenlin Meng and
                  Stefano Ermon},
  title        = {Denoising Diffusion Implicit Models},
  booktitle    = {{ICLR}},
  publisher    = {OpenReview.net},
  year         = {2021}
}

@inproceedings{smplx,
  author       = {Georgios Pavlakos and
                  Vasileios Choutas and
                  Nima Ghorbani and
                  Timo Bolkart and
                  Ahmed A. A. Osman and
                  Dimitrios Tzionas and
                  Michael J. Black},
  title        = {Expressive Body Capture: 3D Hands, Face, and Body From a Single Image},
  booktitle    = {{CVPR}},
  pages        = {10975--10985},
  publisher    = {Computer Vision Foundation / {IEEE}},
  year         = {2019}
}

@inproceedings{wav2vec,
  author       = {Alexei Baevski and
                  Yuhao Zhou and
                  Abdelrahman Mohamed and
                  Michael Auli},
  title        = {wav2vec 2.0: {A} Framework for Self-Supervised Learning of Speech
                  Representations},
  booktitle    = {NeurIPS},
  year         = {2020}
}

@inproceedings{hamer,
  author       = {Georgios Pavlakos and
                  Dandan Shan and
                  Ilija Radosavovic and
                  Angjoo Kanazawa and
                  David Fouhey and
                  Jitendra Malik},
  title        = {Reconstructing Hands in 3D with Transformers},
  booktitle    = {{CVPR}},
  pages        = {9826--9836},
  publisher    = {{IEEE}},
  year         = {2024}
}

@inproceedings{PyMAF,
  author       = {Hongwen Zhang and
                  Yating Tian and
                  Xinchi Zhou and
                  Wanli Ouyang and
                  Yebin Liu and
                  Limin Wang and
                  Zhenan Sun},
  title        = {PyMAF: 3D Human Pose and Shape Regression with Pyramidal Mesh Alignment
                  Feedback Loop},
  booktitle    = {{ICCV}},
  pages        = {11426--11436},
  publisher    = {{IEEE}},
  year         = {2021}
}

@inproceedings{probtalk,
  author       = {Yifei Liu and
                  Qiong Cao and
                  Yandong Wen and
                  Huaiguang Jiang and
                  Changxing Ding},
  title        = {Towards Variable and Coordinated Holistic Co-Speech Motion Generation},
  booktitle    = {{CVPR}},
  pages        = {1566--1576},
  publisher    = {{IEEE}},
  year         = {2024}
}

@article{wang2022vpu,
  title={VPU: A video-based point cloud upsampling framework},
  author={Wang, Kaisiyuan and Sheng, Lu and Gu, Shuhang and Xu, Dong},
  journal={IEEE Transactions on Image Processing},
  volume={31},
  pages={4062--4075},
  year={2022},
  publisher={IEEE}
}

@inproceedings{talkshow,
  author       = {Hongwei Yi and
                  Hualin Liang and
                  Yifei Liu and
                  Qiong Cao and
                  Yandong Wen and
                  Timo Bolkart and
                  Dacheng Tao and
                  Michael J. Black},
  title        = {Generating Holistic 3D Human Motion from Speech},
  booktitle    = {{CVPR}},
  pages        = {469--480},
  publisher    = {{IEEE}},
  year         = {2023}
}

@inproceedings{diffstyle,
  author       = {Sicheng Yang and
                  Zhiyong Wu and
                  Minglei Li and
                  Zhensong Zhang and
                  Lei Hao and
                  Weihong Bao and
                  Ming Cheng and
                  Long Xiao},
  title        = {DiffuseStyleGesture: Stylized Audio-Driven Co-Speech Gesture Generation
                  with Diffusion Models},
  booktitle    = {{IJCAI}},
  pages        = {5860--5868},
  publisher    = {ijcai.org},
  year         = {2023}
}

@inproceedings{SIGGesture,
  author       = {Qingrong Cheng and
                  Xu Li and
                  Xinghui Fu},
  title        = {SIGGesture: Generalized Co-Speech Gesture Synthesis via Semantic Injection
                  with Large-Scale Pre-Training Diffusion Models},
  booktitle    = {{SIGGRAPH} Asia},
  pages        = {133:1--133:11},
  publisher    = {{ACM}},
  year         = {2024}
}

@article{sege,
  author       = {Zeyi Zhang and
                  Tenglong Ao and
                  Yuyao Zhang and
                  Qingzhe Gao and
                  Chuan Lin and
                  Baoquan Chen and
                  Libin Liu},
  title        = {Semantic Gesticulator: Semantics-Aware Co-Speech Gesture Synthesis},
  journal      = {{ACM} Trans. Graph.},
  volume       = {43},
  number       = {4},
  pages        = {136:1--136:17},
  year         = {2024}
}

@inproceedings{HA2G_ted,
  author       = {Xian Liu and
                  Qianyi Wu and
                  Hang Zhou and
                  Yinghao Xu and
                  Rui Qian and
                  Xinyi Lin and
                  Xiaowei Zhou and
                  Wayne Wu and
                  Bo Dai and
                  Bolei Zhou},
  title        = {Learning Hierarchical Cross-Modal Association for Co-Speech Gesture
                  Generation},
  booktitle    = {{CVPR}},
  pages        = {10452--10462},
  publisher    = {{IEEE}},
  year         = {2022}
}

@inproceedings{vpgc,
  title={Efficient video portrait reenactment via grid-based codebook},
  author={Wang, Kaisiyuan and Zhou, Hang and Wu, Qianyi and Tang, Jiaxiang and Xu, Zhiliang and Liang, Borong and Hu, Tianshu and Ding, Errui and Liu, Jingtuo and Liu, Ziwei and others},
  booktitle={ACM SIGGRAPH 2023 Conference Proceedings},
  pages={1--9},
  year={2023}
}

@inproceedings{vprq,
  title={Robust video portrait reenactment via personalized representation quantization},
  author={Wang, Kaisiyuan and Liang, Changcheng and Zhou, Hang and Tang, Jiaxiang and Wu, Qianyi and He, Dongliang and Hong, Zhibin and Liu, Jingtuo and Ding, Errui and Liu, Ziwei and others},
  booktitle={Proceedings of the AAAI Conference on Artificial Intelligence},
  volume={37},
  number={2},
  pages={2564--2572},
  year={2023}
}

@inproceedings{beat,
  author       = {Haiyang Liu and
                  Zihao Zhu and
                  Naoya Iwamoto and
                  Yichen Peng and
                  Zhengqing Li and
                  You Zhou and
                  Elif Bozkurt and
                  Bo Zheng},
  title        = {{BEAT:} {A} Large-Scale Semantic and Emotional Multi-modal Dataset
                  for Conversational Gestures Synthesis},
  booktitle    = {{ECCV} {(7)}},
  series       = {Lecture Notes in Computer Science},
  volume       = {13667},
  pages        = {612--630},
  publisher    = {Springer},
  year         = {2022}
}

@inproceedings{EMAGE,
  author       = {Haiyang Liu and
                  Zihao Zhu and
                  Giorgio Becherini and
                  Yichen Peng and
                  Mingyang Su and
                  You Zhou and
                  Xuefei Zhe and
                  Naoya Iwamoto and
                  Bo Zheng and
                  Michael J. Black},
  title        = {{EMAGE:} Towards Unified Holistic Co-Speech Gesture Generation via
                  Expressive Masked Audio Gesture Modeling},
  booktitle    = {{CVPR}},
  pages        = {1144--1154},
  publisher    = {{IEEE}},
  year         = {2024}
}

@inproceedings{DiffSHEG,
  author       = {Junming Chen and
                  Yunfei Liu and
                  Jianan Wang and
                  Ailing Zeng and
                  Yu Li and
                  Qifeng Chen},
  title        = {DiffSHEG: {A} Diffusion-Based Approach for Real-Time Speech-Driven
                  Holistic 3D Expression and Gesture Generation},
  booktitle    = {{CVPR}},
  pages        = {7352--7361},
  publisher    = {{IEEE}},
  year         = {2024}
}

@inproceedings{Audio2Gestures,
  author       = {Jing Li and
                  Di Kang and
                  Wenjie Pei and
                  Xuefei Zhe and
                  Ying Zhang and
                  Zhenyu He and
                  Linchao Bao},
  title        = {Audio2Gestures: Generating Diverse Gestures from Speech Audio with
                  Conditional Variational Autoencoders},
  booktitle    = {{ICCV}},
  pages        = {11273--11282},
  publisher    = {{IEEE}},
  year         = {2021}
}

@inproceedings{SEEG,
  author       = {Yuanzhi Liang and
                  Qianyu Feng and
                  Linchao Zhu and
                  Li Hu and
                  Pan Pan and
                  Yi Yang},
  title        = {{SEEG:} Semantic Energized Co-speech Gesture Generation},
  booktitle    = {{CVPR}},
  pages        = {10463--10472},
  publisher    = {{IEEE}},
  year         = {2022}
}

@inproceedings{DiffGesture,
  author       = {Lingting Zhu and
                  Xian Liu and
                  Xuanyu Liu and
                  Rui Qian and
                  Ziwei Liu and
                  Lequan Yu},
  title        = {Taming Diffusion Models for Audio-Driven Co-Speech Gesture Generation},
  booktitle    = {{CVPR}},
  pages        = {10544--10553},
  publisher    = {{IEEE}},
  year         = {2023}
}

@article{co-speech-Rhythmic,
  author       = {Tenglong Ao and
                  Qingzhe Gao and
                  Yuke Lou and
                  Baoquan Chen and
                  Libin Liu},
  title        = {Rhythmic Gesticulator: Rhythm-Aware Co-Speech Gesture Synthesis with
                  Hierarchical Neural Embeddings},
  journal      = {{ACM} Trans. Graph.},
  volume       = {41},
  number       = {6},
  pages        = {209:1--209:19},
  year         = {2022}
}

@inproceedings{co-speech-rule1,
  author       = {Justine Cassell and
                  Hannes H{\"{o}}gni Vilhj{\'{a}}lmsson and
                  Timothy W. Bickmore},
  title        = {{BEAT:} the Behavior Expression Animation Toolkit},
  booktitle    = {{SIGGRAPH}},
  pages        = {477--486},
  publisher    = {{ACM}},
  year         = {2001}
}

@article{co-speech-rule2,
  author       = {Sergey Levine and
                  Philipp Kr{\"{a}}henb{\"{u}}hl and
                  Sebastian Thrun and
                  Vladlen Koltun},
  title        = {Gesture controllers},
  journal      = {{ACM} Trans. Graph.},
  volume       = {29},
  number       = {4},
  pages        = {124:1--124:11},
  year         = {2010}
}

@article{co-speech-rule3,
  author       = {Stefan Kopp and
                  Ipke Wachsmuth},
  title        = {Synthesizing multimodal utterances for conversational agents},
  journal      = {Comput. Animat. Virtual Worlds},
  volume       = {15},
  number       = {1},
  pages        = {39--52},
  year         = {2004}
}

@article{co-speech-rule4,
  author       = {Petra Wagner and
                  Zofia Malisz and
                  Stefan Kopp},
  title        = {Gesture and speech in interaction: An overview},
  journal      = {Speech Commun.},
  volume       = {57},
  pages        = {209--232},
  year         = {2014}
}

@inproceedings{co-speech-rule5,
  author       = {Stacy Marsella and
                  Yuyu Xu and
                  Margaux Lhommet and
                  Andrew W. Feng and
                  Stefan Scherer and
                  Ari Shapiro},
  title        = {Virtual character performance from speech},
  booktitle    = {Symposium on Computer Animation},
  pages        = {25--35},
  publisher    = {{ACM}},
  year         = {2013}
}

@article{co-speech-1,
  author       = {Youngwoo Yoon and
                  Bok Cha and
                  Joo{-}Haeng Lee and
                  Minsu Jang and
                  Jaeyeon Lee and
                  Jaehong Kim and
                  Geehyuk Lee},
  title        = {Speech gesture generation from the trimodal context of text, audio,
                  and speaker identity},
  journal      = {{ACM} Trans. Graph.},
  volume       = {39},
  number       = {6},
  pages        = {222:1--222:16},
  year         = {2020}
}

@article{co-speech-2,
  author       = {Simon Alexanderson and
                  Gustav Eje Henter and
                  Taras Kucherenko and
                  Jonas Beskow},
  title        = {Style-Controllable Speech-Driven Gesture Synthesis Using Normalising
                  Flows},
  journal      = {Comput. Graph. Forum},
  volume       = {39},
  number       = {2},
  pages        = {487--496},
  year         = {2020}
}

@inproceedings{co-speech-3,
  author       = {Taras Kucherenko and
                  Patrik Jonell and
                  Sanne van Waveren and
                  Gustav Eje Henter and
                  Simon Alexandersson and
                  Iolanda Leite and
                  Hedvig Kjellstr{\"{o}}m},
  title        = {Gesticulator: {A} framework for semantically-aware speech-driven gesture
                  generation},
  booktitle    = {{ICMI}},
  pages        = {242--250},
  publisher    = {{ACM}},
  year         = {2020}
}

@inproceedings{co-speech-4-lstm,
  author       = {Dai Hasegawa and
                  Naoshi Kaneko and
                  Shinichi Shirakawa and
                  Hiroshi Sakuta and
                  Kazuhiko Sumi},
  title        = {Evaluation of Speech-to-Gesture Generation Using Bi-Directional {LSTM}
                  Network},
  booktitle    = {{IVA}},
  pages        = {79--86},
  publisher    = {{ACM}},
  year         = {2018}
}

@inproceedings{co-speech-5,
  author       = {Mingyang Sun and
                  Mengchen Zhao and
                  Yaqing Hou and
                  Minglei Li and
                  Huang Xu and
                  Songcen Xu and
                  Jianye Hao},
  title        = {Co-speech Gesture Synthesis by Reinforcement Learning with Contrastive
                  Pretrained Rewards},
  booktitle    = {{CVPR}},
  pages        = {2331--2340},
  publisher    = {{IEEE}},
  year         = {2023}
}

@inproceedings{DisCo,
  author       = {Haiyang Liu and
                  Naoya Iwamoto and
                  Zihao Zhu and
                  Zhengqing Li and
                  You Zhou and
                  Elif Bozkurt and
                  Bo Zheng},
  title        = {DisCo: Disentangled Implicit Content and Rhythm Learning for Diverse
                  Co-Speech Gestures Synthesis},
  booktitle    = {{ACM} Multimedia},
  pages        = {3764--3773},
  publisher    = {{ACM}},
  year         = {2022}
}

@inproceedings{resyncer,
  title={Resyncer: Rewiring Style-based Generator for Unified Audio-Visually Synced Facial Performer},
  author={Guan, Jiazhi and Xu, Zhiliang and Zhou, Hang and Wang, Kaisiyuan and He, Shengyi and Zhang, Zhanwang and Liang, Borong and Feng, Haocheng and Ding, Errui and Liu, Jingtuo and others},
  booktitle={European Conference on Computer Vision},
  pages={348--367},
  year={2025},
  organization={Springer}
}

@article{attention,
  title={Attention is all you need},
  author={Vaswani, A},
  journal={Advances in Neural Information Processing Systems},
  year={2017}
}

@article{learndance,
  title={Learn to dance with aist++: Music conditioned 3d dance generation},
  author={Li, Ruilong and Yang, Shan and Ross, David A and Kanazawa, Angjoo},
  journal={arXiv preprint arXiv:2101.08779},
  volume={2},
  number={3},
  year={2021},
  publisher={eprint}
}

@inproceedings{liu2022learning,
  title={Learning hierarchical cross-modal association for co-speech gesture generation},
  author={Liu, Xian and Wu, Qianyi and Zhou, Hang and Xu, Yinghao and Qian, Rui and Lin, Xinyi and Zhou, Xiaowei and Wu, Wayne and Dai, Bo and Zhou, Bolei},
  booktitle={Proceedings of the IEEE/CVF Conference on Computer Vision and Pattern Recognition},
  pages={10462--10472},
  year={2022}
}

@misc{tevet2022humanmotiondiffusionmodel,
      title={Human Motion Diffusion Model}, 
      author={Guy Tevet and Sigal Raab and Brian Gordon and Yonatan Shafir and Daniel Cohen-Or and Amit H. Bermano},
      year={2022},
      eprint={2209.14916},
      archivePrefix={arXiv},
      primaryClass={cs.CV},
      url={https://arxiv.org/abs/2209.14916}, 
}

@inproceedings{ldm,
  title={High-resolution image synthesis with latent diffusion models},
  author={Rombach, Robin and Blattmann, Andreas and Lorenz, Dominik and Esser, Patrick and Ommer, Bj{\"o}rn},
  booktitle={Proceedings of the IEEE/CVF conference on computer vision and pattern recognition},
  pages={10684--10695},
  year={2022}
}

@inproceedings{dwpose,
  title={Effective whole-body pose estimation with two-stages distillation},
  author={Yang, Zhendong and Zeng, Ailing and Yuan, Chun and Li, Yu},
  booktitle={Proceedings of the IEEE/CVF International Conference on Computer Vision},
  pages={4210--4220},
  year={2023}
}

@inproceedings{ReMoDiffuse,
  author       = {Mingyuan Zhang and
                  Xinying Guo and
                  Liang Pan and
                  Zhongang Cai and
                  Fangzhou Hong and
                  Huirong Li and
                  Lei Yang and
                  Ziwei Liu},
  title        = {ReMoDiffuse: Retrieval-Augmented Motion Diffusion Model},
  booktitle    = {{ICCV}},
  pages        = {364--373},
  publisher    = {{IEEE}},
  year         = {2023}
}

@article{hands2minds,
  title={From hands to minds: Gestures promote understanding},
  author={Kang, Seokmin and Tversky, Barbara},
  journal={Cognitive Research: Principles and Implications},
  volume={1},
  pages={1--15},
  year={2016},
  publisher={Springer}
}

@article{learn-to-count,
  title={The function of gesture in learning to count: More than keeping track},
  author={Alibali, Martha Wagner and DiRusso, Alyssa A},
  journal={Cognitive development},
  volume={14},
  number={1},
  pages={37--56},
  year={1999},
  publisher={Elsevier}
}

@inproceedings{animateanyone,
  author       = {Li Hu},
  title        = {Animate Anyone: Consistent and Controllable Image-to-Video Synthesis
                  for Character Animation},
  booktitle    = {{CVPR}},
  pages        = {8153--8163},
  publisher    = {{IEEE}},
  year         = {2024}
}

@inproceedings{MDT-A2G,
  author       = {Xiaofeng Mao and
                  Zhengkai Jiang and
                  Qilin Wang and
                  Chencan Fu and
                  Jiangning Zhang and
                  Jiafu Wu and
                  Yabiao Wang and
                  Chengjie Wang and
                  Wei Li and
                  Mingmin Chi},
  title        = {{MDT-A2G:} Exploring Masked Diffusion Transformers for Co-Speech Gesture
                  Generation},
  booktitle    = {{ACM} Multimedia},
  pages        = {3266--3274},
  publisher    = {{ACM}},
  year         = {2024}
}

@inproceedings{MambaTalk,
  author       = {Zunnan Xu and
                  Yukang Lin and
                  Haonan Han and
                  Sicheng Yang and
                  Ronghui Li and
                  Yachao Zhang and
                  Xiu Li},
  title        = {MambaTalk: Efficient Holistic Gesture Synthesis with Selective State
                  Space Models},
  booktitle    = {NeurIPS},
  year         = {2024}
}

@inproceedings{SGToolkit,
  author       = {Youngwoo Yoon and
                  Keunwoo Park and
                  Minsu Jang and
                  Jaehong Kim and
                  Geehyuk Lee},
  title        = {SGToolkit: An Interactive Gesture Authoring Toolkit for Embodied Conversational
                  Agents},
  booktitle    = {{UIST}},
  pages        = {826--840},
  publisher    = {{ACM}},
  year         = {2021}
}
}

\clearpage 
\appendix
\section*{Appendix}
\section{Dataset}
\subsection{Dataset Details}
\begin{figure}[h]
    \centering
    \includegraphics[width=0.8\linewidth]{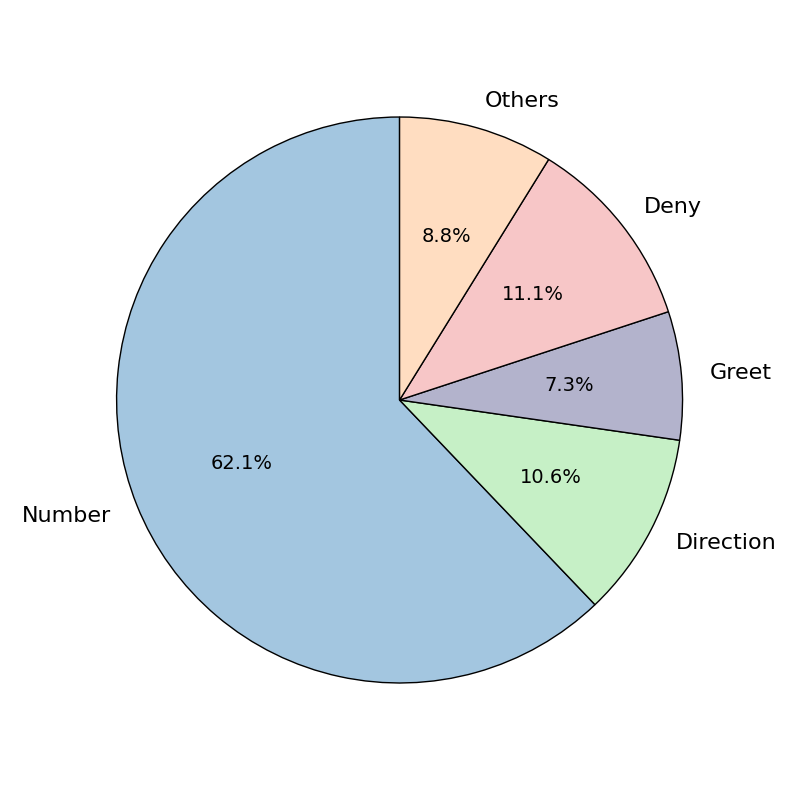}
    \caption{The distribution of the number of different semantic gestures.}
    \label{fig:action}
\end{figure}

Our curated Streamer dataset includes 18 categories of gestures with specific semantics, such as numbers, directions, Greet, and Deny. 
In addition, some parallel sentences in the dataset also contain specific semantic gestures.
The details of the semantic gestures are shown in Table~\ref{tab:data_type}, and the distribution of different semantic gestures is illustrated in Fig.~\ref{fig:action}.

\begin{table}[h]
    \centering
    \caption{The Types of Semantic Gestures in the Streamer Dataset}
    \renewcommand{\arraystretch}{1.3} 
    \setlength{\tabcolsep}{8pt} 
    \small 
    \resizebox{0.5\textwidth}{!}{ 
        \begin{tabular}{|>{\columncolor[HTML]{D3D3D3}}c|p{3cm}|p{4.8cm}|} 
        \hline
        \rowcolor[HTML]{D3D3D3} 
        \textbf{Types} & \textbf{Contents} & \textbf{Examples} \\ \hline
        Number         & 1-10  & We'll give you the reduction of \textbf{four} hundred yuan \\ \hline
        Direction      & Upper, Lower, Upper left, Lower left, Upper right, Lower right & Click the \textbf{lower right} corner to add it to your collection now. \\ \hline
        Greet        & Hello (Hi)  & \textbf{Hello} and welcome to our live room. \\ \hline
        Deny         & Don’t (Doesn't, Not) & It’s really \textbf{not} expensive. \\
        \hline
        Others       & Parallel sentences & Whether it is \textbf{boys, girls, the elderly, or children}, can be used. \\
        \hline
        \end{tabular}
    }
    \label{tab:data_type}
\end{table}

\begin{figure*}[t]
    \centering
    \includegraphics[width=\linewidth]{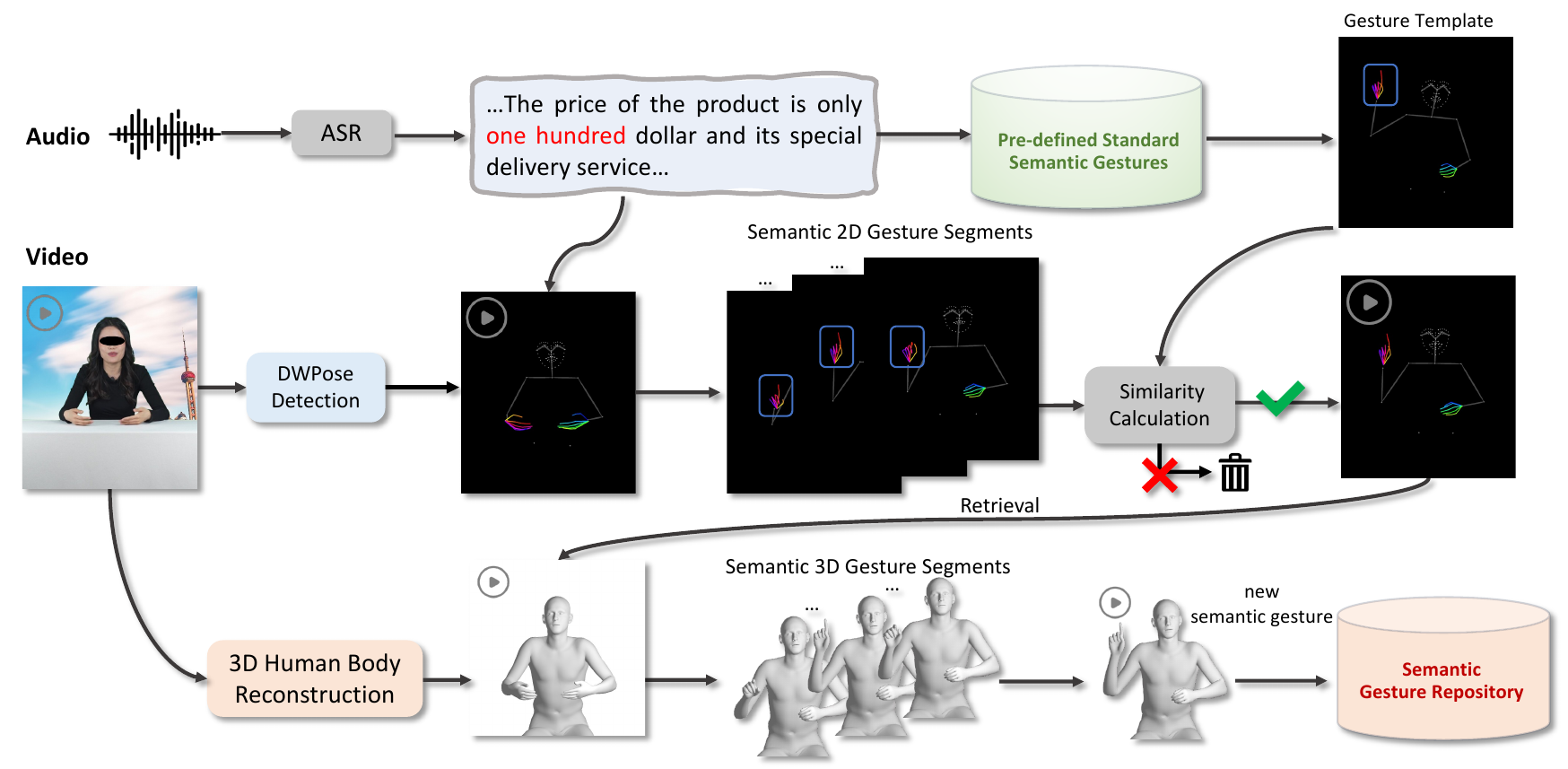}
    \caption{Automatic Annotation Pipeline for Personalized Semantic Gesture Repository.}
    \label{fig:anno_pipeline}
\end{figure*}

\begin{figure}[h]
    \centering
    \includegraphics[width=0.8\linewidth]{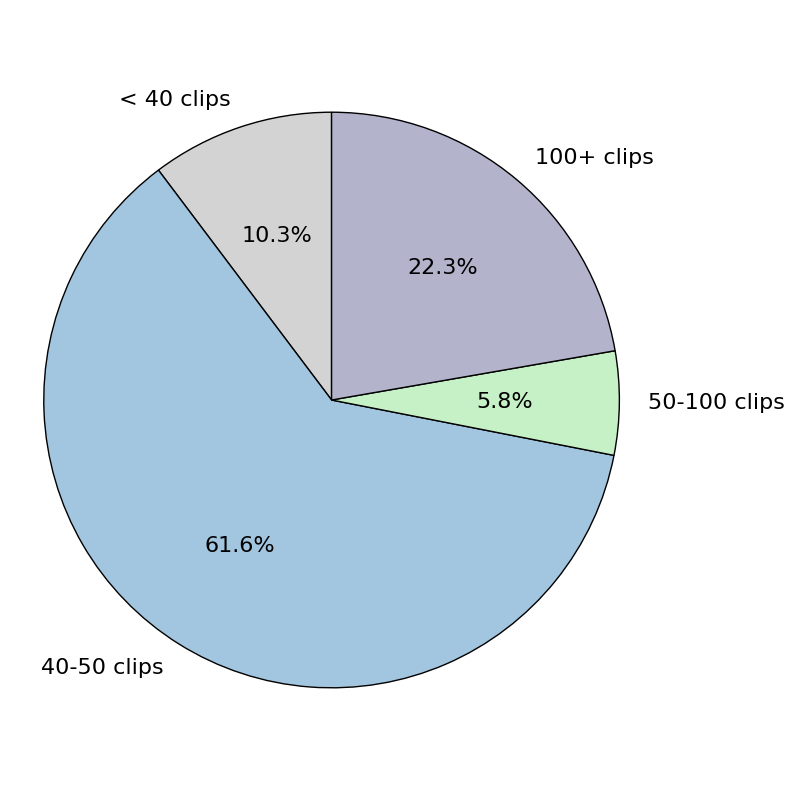}
    \caption{The distribution of the number of short clips for different anchors.}
    \label{fig:distribution}
\end{figure}

Our dataset includes 281 anchor actors, with a total video duration of about 58 hours. 
Each video is split into 10-second short clips, with a frame rate of 25 fps and an audio sampling rate of 22 kHz. 
The processed dataset contains 20,969 short clips, and the distribution of clip numbers for different anchors is shown in Fig.~\ref{fig:distribution}. The majority of actors have between 40 and 50 short clips.

After obtaining the clips, we follow SHOW~\cite{talkshow} to reconstruct the SMPL-X~\cite{smplx} parameters from monocular videos. Our work focuses on body and gesture generation, so we do not reconstruct facial expression coefficients. Additionally, while SHOW uses PyMAF-X~\cite{PyMAF} to initialize hand poses, we replace PyMAF-X with the more powerful hamer~\cite{hamer} to improve the quality of initial hand reconstruction.

\subsection{Automatic Annotation Approach}

Our framework is also equipped with an automatic semantic gesture annotation procedure (shown in Fig.~\ref{fig:anno_pipeline}) that can extract semantic gestures from a video of each target person. 
Specifically, we first prepare 18 pre-defined standard semantic gestures in the form of 2D skeletons by using DWPose~\cite{dwpose} detection, which can be regarded as semantic gesture templates. Given a target video, we first use automatic-speech-recognition (ASR) technique to convert its audio into text, and then locate the segments aligned with special trigger words (e.g., "first", "left" , and "hello"). By performing DWPose detection on these segments, we are able to calculate the similarity between the detected gestures and the corresponding pre-defined gesture template. It's worth noting that semantic gestures may not be triggered every time when a person speaks the same word, thus such a process is necessary to filter relevant segments for personalized gesture collection. Once a segment exhibits high consistency with the gesture template, we continue to perform the 3D human body reconstruction process, and the reconstructed result is accepted as a standard example of the corresponding gesture. Normally, we collect at least two examples for each gesture unless a specific gesture is quite rare or missing in the captured data.

\section{More Experimental Analysis}
\begin{table}[t]
\centering
\caption{The statistical results of the user study}
\label{tab:user_study}
\resizebox{0.48\textwidth}{!}{ 
\begin{tabular}{lccc}
\hline
System 
& \textbf{Naturalness$\uparrow$} & \textbf{Rhythm $\uparrow$} & \textbf{Semantic$\uparrow$} \\ \hline
TalkSHOW        & 2.53 & 2.67 & 2.36      \\
Probtalk        & 2.71 & 2.89 & 2.80      \\
DSG             & 2.89 & 3.05 & 2.82      \\
Ours            & \textbf{3.82} & \textbf{3.84} &   \textbf{4.29}
          \\ \hline
\end{tabular}
}
\label{table:user}
\end{table}

\subsection{Human Evaluation.}
We perform a user study including 17 examples from the Streamer dataset and 20 volunteers. 
For each example, 4 gesture sequences generated by our method and other SOTAs, accompanied by audio, are displayed in a random order.
Each participant is asked to score the generated gestures based on the following three aspects:
(a) \textbf{Naturalness:} Whether the generated gestures are natural and smooth;
(b) \textbf{Rhythm:} Whether the rhythm of the generated gestures aligns with the audio rhythm;
(c) \textbf{Semantics:} Whether the generated motions match the semantic information in the speech.
The scoring range was from 1 to 5, with higher scores indicating better performance. 
The results in Table~\ref{table:user}, demonstrate that our method significantly outperforms the comparison methods over all three metrics, highlighting the superiority of the gesture modeling ability of our system.

\subsection{More Gesture Editing}
Thanks to our hybrid-modality design, masking training strategy and motion style injection, the proposed GestureHYDRA not only possesses the ability to insert keyframes (i.e. Semantic gesture injection) but also includes other meaningful features, such as motion style transfer, motion inpainting, and motion erase. 
Our model allows the users to perform robust and flexible editing operations on the generated gesture sequence for distinct real-world demands. 

\begin{figure}[h]
    \centering
    \includegraphics[width=0.8\linewidth]{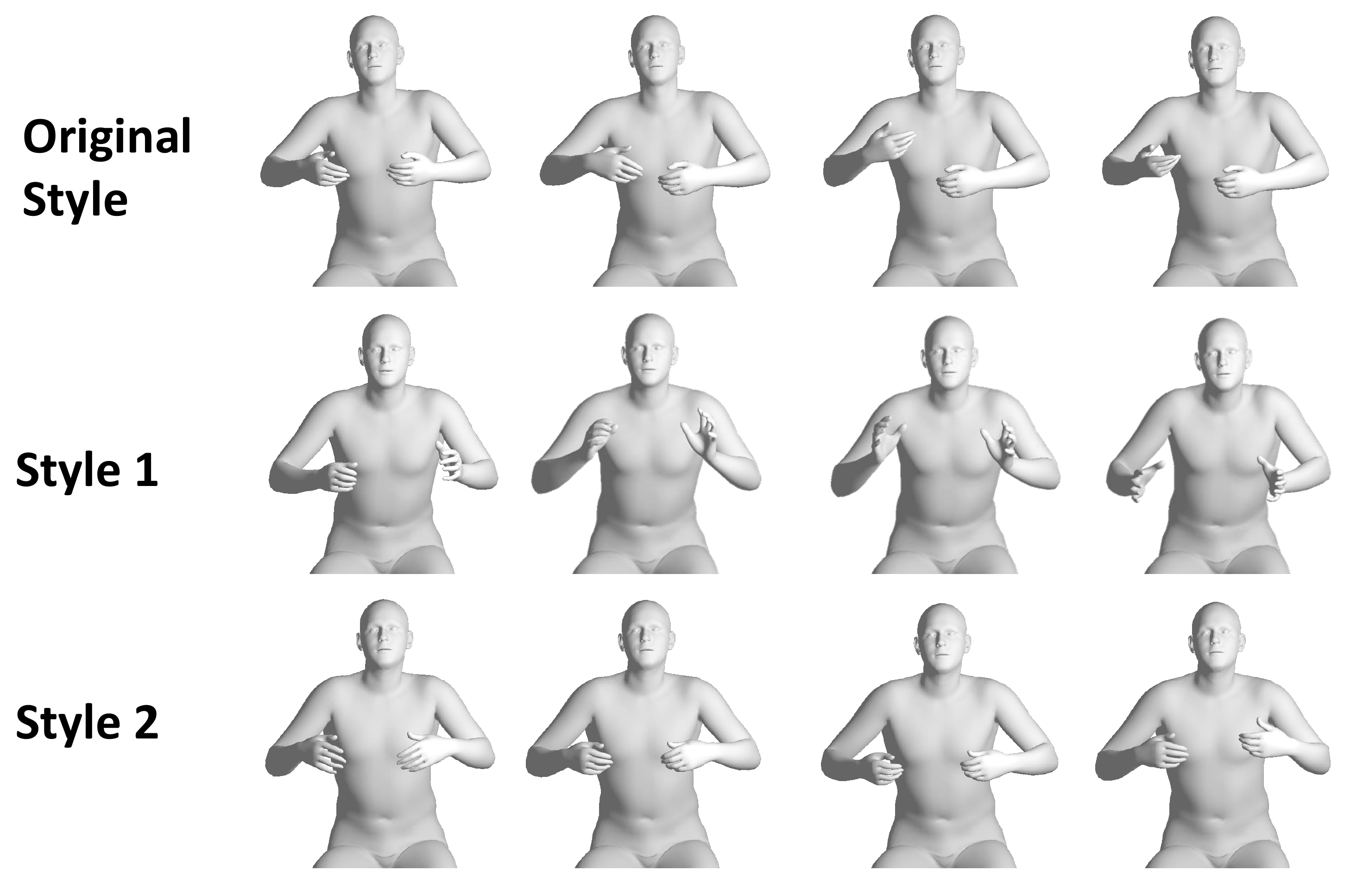}
    \caption{Motions generated in different styles under the same speech.}
    \label{fig:motion_transfer}
\end{figure}

\begin{figure}[h]
    \centering
    \includegraphics[width=0.8\linewidth]{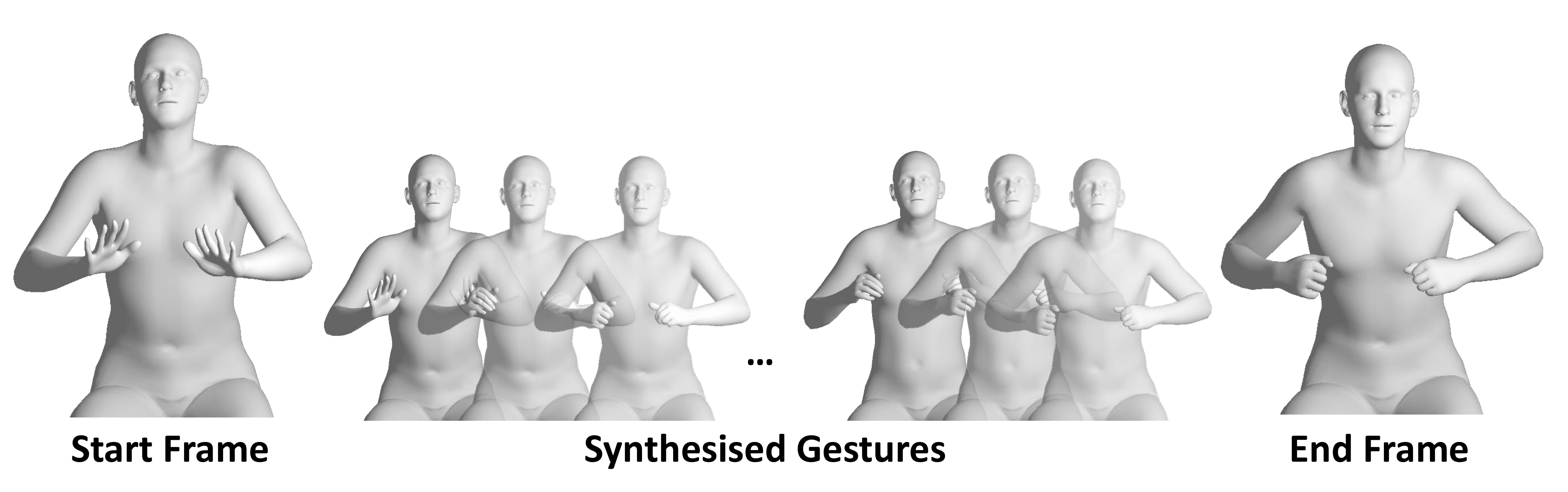}
    \caption{Motion inpainting given start and end frames.}
    \label{fig:motion_inpainting}
\end{figure}

\begin{figure}[h]
    \centering
    \includegraphics[width=0.8\linewidth]{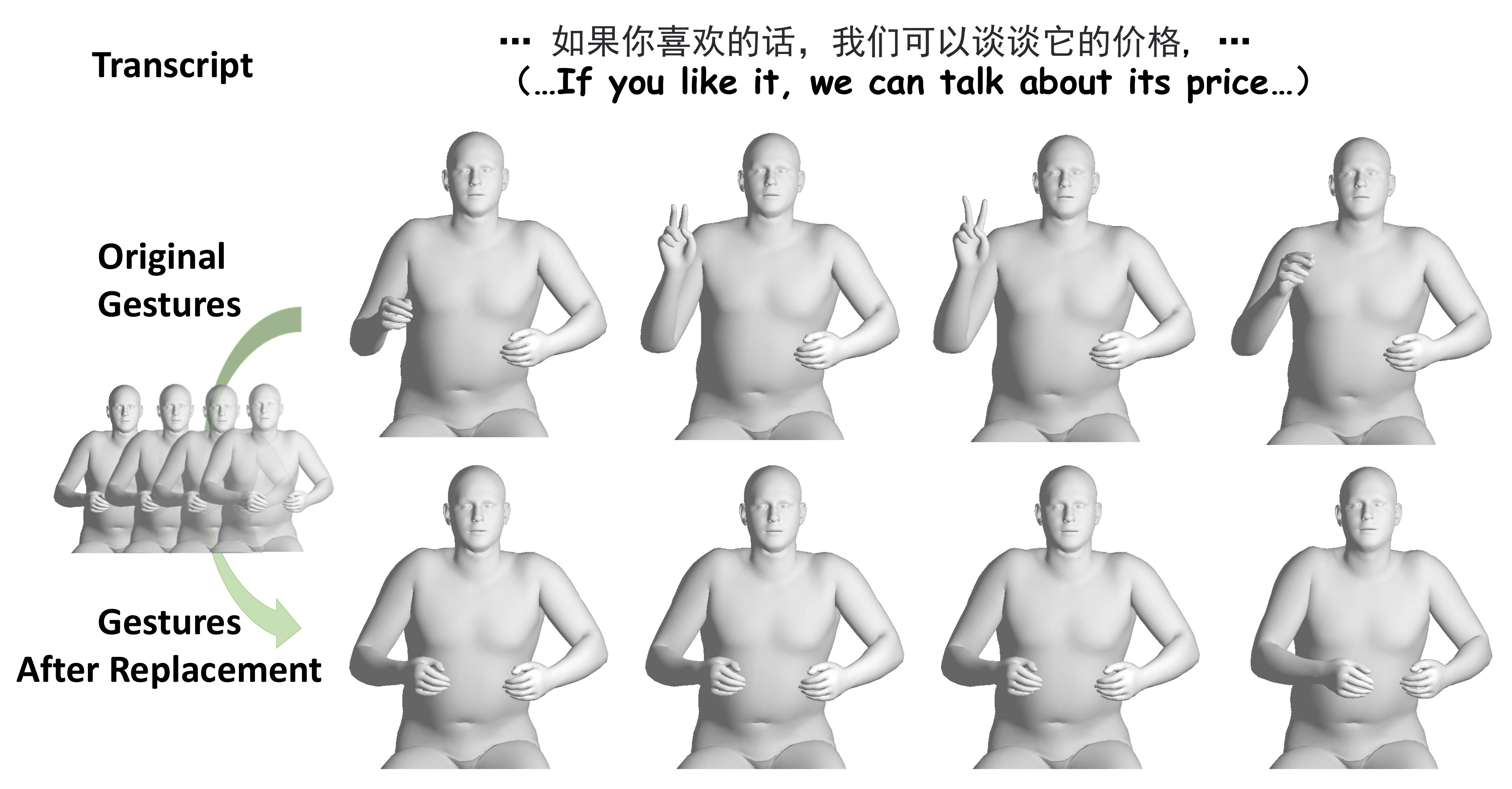}
    \caption{Replace undesired segments in generated motion.}
    \label{fig:motion_erase}
\end{figure}

\textbf{Motion Style Transfer.} Given the same speech input and different style embeddings for various characters, our proposed GestureHYDRA can generate body gestures in different styles, all while synchronizing with the given speech.(See Fig.~\ref{fig:motion_transfer})
\textbf{Motion Inpainting.} 
Given the input speech, starting frame, and ending frame, our proposed GestureHYDRA can automatically generate the motion sequence between the starting and ending frames by inserting them as keyframes at the corresponding positions.
(See Fig.~\ref{fig:motion_inpainting})
\textbf{Motion Segment Replacement.}
Our GestureHYDRA replaces unwanted frame segments in the generated motion with target gesture fragments as keyframes, while also treating the remaining usable frames as keyframes, thus enabling the replacement of undesired generated gestures.(See Fig.~\ref{fig:motion_erase})

Please watch the demo video we provided to view the generated results.

\section{Limitations}
Although our work paves the way for generating gestures with strong semantics, it still has certain limitations: 
1) Our Streamer dataset currently excludes face reconstruction and digital human creation without vivid facial expressions partly degrades user experiences.
2) The designed Retrieval-Augmented Generation strategy relies on correct automatic speech recognition results. Ambiguous or incorrect recognition results will lead to an evident mismatch between speech audio and human gestures.
Our future exploration will advance from these two directions.
\end{document}